\documentclass[11pt]{article}
\input{epsf}
\begin{document}
%

%
\def\tilde{\widetilde}
\def\bar{\overline}
\def\hat{\widehat}
\def\*{\star}
\def\[{\left[}
\def\]{\right]}
\def\({\left(}
\def\){\right)}
\def\zb{{\bar{z} }}
\def\frac#1#2{{#1 \over #2}}
\def\inv#1{{1 \over #1}}
\def\half{{1 \over 2}}
\def\d{\partial}
\def\der#1{{\partial \over \partial #1}}
\def\dd#1#2{{\partial #1 \over \partial #2}}
\def\vev#1{\langle #1 \rangle}
\def\bra#1{{\langle #1 |  }}
\def\ket#1{ | #1 \rangle}
\def\rvac{\hbox{$\vert 0\rangle$}}
\def\lvac{\hbox{$\langle 0 \vert $}}
\def\2pi{\hbox{$2\pi i$}}
\def\e#1{{\rm e}^{^{\textstyle #1}}}
\def\grad#1{\,\nabla\!_{{#1}}\,}
\def\dsl{\raise.15ex\hbox{/}\kern-.57em\partial}
\def\Dsl{\,\raise.15ex\hbox{/}\mkern-.13.5mu D}
\def\comm#1#2{ \BBL\ #1\ ,\ #2 \BBR }
\def\x{\stackrel{\otimes}{,}}
\def\det{ {\rm det}}
\def\tr{{\rm tr}}
%
%
\def\th{\theta}		\def\Th{\Theta}
\def\ga{\gamma}		\def\Ga{\Gamma}
\def\be{\beta}
\def\al{\alpha}
\def\ep{\epsilon}
\def\la{\lambda}	\def\La{\Lambda}
\def\de{\delta}		\def\De{\Delta}
\def\om{\omega}		\def\Om{\Omega}
\def\sig{\sigma}	\def\Sig{\Sigma}
\def\vphi{\varphi}
%
%
\def\CA{{\cal A}}	\def\CB{{\cal B}}	\def\CC{{\cal C}}
\def\CD{{\cal D}}	\def\CE{{\cal E}}	\def\CF{{\cal F}}
\def\CG{{\cal G}}	\def\CH{{\cal H}}	\def\CI{{\cal J}}
\def\CJ{{\cal J}}	\def\CK{{\cal K}}	\def\CL{{\cal L}}
\def\CM{{\cal M}}	\def\CN{{\cal N}}	\def\CO{{\cal O}}
\def\CP{{\cal P}}	\def\CQ{{\cal Q}}	\def\CR{{\cal R}}
\def\CS{{\cal S}}	\def\CT{{\cal T}}	\def\CU{{\cal U}}
\def\CV{{\cal V}}	\def\CW{{\cal W}}	\def\CX{{\cal X}}
\def\CY{{\cal Y}}	\def\CZ{{\cal Z}}
%
%
\font\numbers=cmss12
\font\upright=cmu10 scaled\magstep1
\def\stroke{\vrule height8pt width0.4pt depth-0.1pt}
\def\topfleck{\vrule height8pt width0.5pt depth-5.9pt}
\def\botfleck{\vrule height2pt width0.5pt depth0.1pt}
\def\Zmath{\vcenter{\hbox{\numbers\rlap{\rlap{Z}\kern
		0.8pt\topfleck}\kern
		2.2pt \rlap Z\kern 6pt\botfleck\kern 1pt}}}
\def\Qmath{\vcenter{\hbox{\upright\rlap{\rlap{Q}\kern
                   3.8pt\stroke}\phantom{Q}}}}
\def\Nmath{\vcenter{\hbox{\upright\rlap{I}\kern 1.7pt N}}}
\def\Cmath{\vcenter{\hbox{\upright\rlap{\rlap{C}\kern
                   3.8pt\stroke}\phantom{C}}}}
\def\Rmath{\vcenter{\hbox{\upright\rlap{I}\kern 1.7pt R}}}
\def\Z{\ifmmode\Zmath\else$\Zmath$\fi}
\def\Q{\ifmmode\Qmath\else$\Qmath$\fi}
\def\N{\ifmmode\Nmath\else$\Nmath$\fi}
\def\C{\ifmmode\Cmath\else$\Cmath$\fi}
\def\R{\ifmmode\Rmath\else$\Rmath$\fi}
\def\cadremath#1{\vbox{\hrule\hbox{\vrule\kern8pt\vbox{\kern8pt
			\hbox{$\displaystyle #1$}\kern8pt} 
			\kern8pt\vrule}\hrule}}
\def\proof{\noindent {\underline {Proof}.} }
\def\cqfd{ {\hfill{$\Box$}} }
\def\square{ {\hfill \vrule height6pt width6pt depth1pt} } 
%
%
\def\debut{ \begin{eqnarray} }
\def\fin{ \end{eqnarray} }
\def\non{ \nonumber }
%

%
%
\rightline{SPhT-00-093}
\vskip 1cm
\centerline{\LARGE TURBULENCE FOR (AND BY) AMATEURS}
\bigskip
\vskip 1cm
\vskip1cm
\centerline{\large  Denis Bernard
\footnote[2]{Member of the CNRS; dbernard@spht.saclay.cea.fr} }
\centerline{Service de Physique Th\'eorique de Saclay
\footnote[3]{\it Laboratoire de la Direction des Sciences de la
Mati\`ere du Commisariat \`a l'Energie Atomique.}}
\centerline{F-91191, Gif-sur-Yvette, France.}
\vskip2cm
Abstract \footnote[4]{Series of lectures presented at S\`ete and Saclay
during the summer 2000.}.\\
Series of lectures on statistical turbulence written 
for amateurs but not experts. Elementary aspects and problems of
turbulence in two and three dimensional Navier-Stokes equation
are introduced. A few properties of scalar turbulence and transport 
phenomena in turbulent flows are described. Kraichnan's model of
passive advection is discussed a bit more precisely.\\
{\bf Part 1: Approaching turbulent flows.} 
	Navier-Stokes equation. Cascades and Kolmogorov theory. 
	Modeling statistical turbulence.
	Correlation functions and scaling.\\
{\bf Part 2: Deeper in turbulent flows.} 
	Turbulence in two dimensions. Dissipation and dissipative anomalies. 
	Fokker-Planck equations. Multifractal models.\\
{\bf Part 3: Scalar turbulence.} 
	Transport and Lagrangian trajectories. 
	Kraichnan's passive scalar model.
	Anomalous scalings and universality.\\
{\bf Part 4: Lagrangian trajectories.}	
	Richardson's law.
	Lagrangian flows in Kraichnan's model. Slow modes.
	Breakdown of Lagrangian flows. Batchelor limit.
	Generalized Lagrangian flows and trajectory bundles.\\
{\bf Part 5: Burgers turbulence.} Shocks. Bifractality.

\vfill
\newpage
%
%
%
%

These lectures are meant as an introduction to a few aspects
and problems of statistical modeling of fully developed
turbulence. As such they are clearly not appropriate for experts
but they perhaps will be more useful for amateurs. 

These series of lectures are essentially divided in  two parts.
The first  part presents quite standard materials on fully
developed turbulence in Navier-Stokes equation: energy cascade,
Kolmogorov theory and deviations, etc... The second part deals
with the passive scalar problem which aims at describing transport 
phenomena in turbulent flows. This problem provides a pleasant laboratory
for studying turbulence in which understandable questions may be asked
and sometimes also answered. It shows intermittency phenomena.
The mechanism at the origin of this behavior relies on global modes,
so-called zero-modes, which are solutions of the inviscid equation
of motion of the effective field theory. Their existence ensures
universality of the intermittency. A similar mechanism may hold 
in Navier-Stokes turbulence.
This problem also points towards a few properties of
Lagrangian trajectories which are at the origin of intermittency phenomena. 
Namely, the existence of global conserved quantities preserved by
the Lagrangian trajectories and the breakdown of the Lagrangian flow
as illustrated by Richardson's law. These properties
are expected to be universal in turbulence. 

\bigskip

\section{Approaching turbulent flows.}

\subsection{Navier-Stokes equation and weak solutions.}
Navier-Stokes and Euler equations, \\
Invariance, dimensional analysis and Reynolds number.\\
Existence of solutions and weak solutions.\\

Since Navier and Stokes fluid motions are suspected
to be governed by the Navier-Stokes equation which is a differential 
equation for the velocity field $u(x,t)$ at point $x$ and time $t$:
\debut
\d_t u + \(u\cdot \nabla\) u - \nu \nabla^2 u = \inv{\rho}\(f - \nabla p\)
\label{nsequa}
\fin
with $\nu$ the viscosity of the fluid, 
$\nu\simeq 10^{-2} {\rm cm}^2{\rm s}^{-1}$ for the water,
$\rho$ its density, $p$ the pressure and $f$ the external force. 
This equation has to be supplemented by the continuity equation,
$\d_t \rho + \nabla\cdot(\rho u)=0$, and the fluid equation of state, $f(p,\rho)=0$. 
In most cases one considers incompressible fluids for which 
the density $\rho$ is constant  in time and position. 
We shall set $\rho=1$ by convention. 
The continuity equation then becomes the incompressibility condition:
\debut
\nabla\cdot u = 0 \label{incomp}
\fin
The pressure $p$ is then not an independent variable since
choosing the external force to satisfy $\nabla\cdot f=0$
and taking the divergence of the Navier-Stokes equation (\ref{nsequa})
gives:
\debut
\nabla^2 p = - (\nabla^j u^k) \,(\nabla_k u_j)
= -(\nabla^j \nabla_k)\, (u^ku_j)\non
\fin
It allows to compute the pressure in terms of the velocity profile.
However, the Navier-Stokes equation is non-local
if it is expressed only in terms of the velocity.

The Euler equation is the inviscid limit $\nu\to 0$ of
the Navier-Stokes equation in which the viscous term $-\nu \nabla^2u$
has been discarded. It possesses a nice geometrical interpretation
as geodesic flow on the group of volume preserving diffeomorphisms.

Once the density has been set to one, the dimensions of the fields
are:
$$
[u]=\frac{\rm length}{\rm time}\quad,\quad
[p]=\frac{{\rm length}^2}{{\rm time}^2} \quad,\quad
[f]=\frac{\rm length}{{\rm time}^2}\quad,\quad 
[\nu] = \frac{{\rm length}^2}{\rm time}
$$
The Navier-Stokes equation is invariant under rescalings,
\debut
u(x,t) &\to& \hat u(x,t) = l\, \tau^{-1}\ u(x/l,t/\tau) \non\\
p(x,t) &\to& \hat p(x,t) = l^2\,\tau^{-2}\ p(x/l,t/\tau) \non\\
f(x,t) &\to& \hat f(x,t) = l\,\tau^{-2}\ f(x/l,t/\tau) \non\\
\nu &\to& \hat \nu = l^2\,\tau^{-1}\ \nu \non
\fin
so that if $u,\ p,\ f$ are solutions of the Navier-Stokes equation
with viscosity $\nu$ so are $\hat u,\ \hat p,\ \hat f$ with
a viscosity $\hat \nu$.
It is thus convenient to introduce the dimensionless
Reynolds number
\debut
\CR e = \frac{(\de_Lu)\ L }{\nu} \label{reynolds}
\fin
where $\de_L u$ is a typical value of the velocity difference
on a scale $L$, a typical length of the system.
Note that this is a scale dependent definition.
For low Reynolds number, $\CR e\ll 1$ the fluid motion is
regular and laminar. For intermediate Reynolds number
$\CR e$ of order $\sim 1$ to $\sim 10^2$ complicated flows
are observed with, depending on the precise set-up,
some of the symmetries permitted by the
equations of motion and the boundary conditions broken.
At higher Reynolds number $\CR e\gg 1$, 
say of order $\sim 10^4$ or more,
the fluid motion shows an apparent spatial disorder, which
seems to be characterized by the proliferation of eddies of
all scales, but with statistical restoration of the symmetries: 
statistical isotropy and translation invariance.
The limit of infinite, ie. very large, Reynolds number is
called {\it fully developed turbulence}. 
Formally it corresponds to the inviscid limit $\nu\to 0$ at
typical length and typical velocity fixed.

Fully developed turbulence is carried by irregular
solutions of the Navier-Stokes equation.
The notion of weak solutions was introduced by Leray. 
It consists in considering
solutions in the sense of distributions. Namely,
$u(x,t)$ is said to be a weak solution of the Navier-Stokes 
equation whenever it satisfies
\debut
\int \[ u^j\,\d_t + u^ju^k\, \nabla^k +
\nu u^j\,\nabla^2 + f^j\]\,\varphi^j=0
\quad,\quad
\int u^j\, \nabla^j\psi =0\non
\fin
for $\varphi^j$, with $\nabla\cdot\varphi=0$, and $\psi$
smooth functions with compact support. Note that
no derivative of the velocity field is taken in
that definition. This allows to consider weak solutions
for which the velocity fields possess singularities in its derivatives.
Weak solutions to the 3d Navier-Stokes equation are known to exist.

The existence of smooth solutions to the 3d Navier-Stokes or Euler equation
with smooth initial data is still an unsolved problem. It is known
that such solutions exist for short time but their existences for all time
is still unclear, ie. do they blow up at finite time or not?
In 2d the situation is different: global existence of solutions 
to the 2d Navier-Stokes has been proved for a large class of initial data. 
See eg. \cite{temam} for more informations.
Classical textbooks on fluid dynamics and turbulence are eg. \cite{yaglom,frisch}.

\subsection{Cascades and Kolmogorov theory.}
Inertial range and Richardson cascades.\\
Hand-waving Kolmogorov theory and K-ology.\\

The Reynolds number (\ref{reynolds})  may be seen as the ratio of
the non-linear advection term  $(u\cdot \nabla)u$
of the Navier-Stokes equation over the dissipation term $-\nu\nabla^2u$.
So that non-linearity dominates in fully developed turbulence.
The advection term preserves the energy $\CE=\int \frac{u^2}{2}$
whereas the dissipation term does not.
From the Navier-Stokes equation the energy balance is:
\debut
\frac{d}{dt}\int\frac{u^2}{2} = \int\[\, f\cdot u - \nu (\nabla u)^2 \,\]
\label{bilan}
\fin
The first term $f\cdot u$ is the energy injected into the system per 
unit of time while the second term $ \nu (\nabla u)^2$ is the
energy dissipated by unit of time.
The pressure does not produce work in incompressible fluids.

In most practical situations the energy is injected into the turbulent
system at large scales, eg the sizes of the bath in which the fluid moves.
It propagates through all system scales and is then dissipated
at small scales at which the dissipation dominates over the advection. 
The large scale, denoted $L$, at which the energy is injected, 
is usually called the integral scale. It serves
as an infrared cut-off. The small scale, denoted $\eta$, at which
dissipation takes place, is called the dissipative scale.
It serves as an ultraviolet cut-off. The intermediate
domain of scales is called  the {\it inertial range}:
$$
\eta \ll {\rm Inertial \ Range} \ll L
$$

The standard characteristic picture for turbulence
is a constant transfer of energy from large scales to small scales
in the inertial range. 
Following Richardson and Kolmogorov this cascade of energy propagates
through the scales via a cascade of eddies: 
big eddies break into smaller ones. 
This picture leads to `mean field' scaling laws.
Suppose that there is a hierarchy of eddies of smaller and smaller
scale $l_n\simeq l_0 \varsigma^n$, $n=0,1,\cdots$, with $\varsigma<1$ 
the contraction ratio of the eddy size from a generation to the following. 
Let $V_{l_n}$ be the volume occupied by eddies of size ${l_n}$.
Since the density of energy carried by the $n$-th generation of eddies
is $u^2_{l_n}/2$, the total energy accumulated in eddies of
size $l_n$ is $E_{l_n} \simeq\, u^2_{l_n}\, V_{l_n}$ with
$V_{l_n}$ the volume occupied by eddies of size ${l_n}$.
The characteristic turnover time in the $n$-th eddy generation
is $\tau_{l_n}\simeq l_n/u_{l_n}$ and the energy flux at scale
$l\simeq l_n$ is thus:
$$
\pi_l \simeq\, E_l/\tau_l\simeq\, V_l\, u^3_l/l
$$
Demanding that {\it this energy flux is constant}, 
ie. independent of the scale, $\pi_l\simeq {\rm const.}$,
and assuming scale homogeneity in the sense that the volume  
occupied by eddies of size $l$ is independent of their sizes,
$V_l \simeq {\rm const.}$, leads to:
\debut
u_l^3\ \simeq\ \bar \ep\, l \label{magic}
\fin
with $\bar \ep$ the energy flux per unit volume, 
$\pi_l\simeq\, \bar \ep\ V_{l_0}$.
Eq.(\ref{magic}) is the fundamental scaling law of turbulence.
It expresses the fact that the energy transfer per unit of time
is constant through scales in the inertial range.
A more precise version of this law, which is known as the
Kolmogorov $4/5$-law, will be described in the following section.

Note that $\bar \ep$ has dimension 
$$ 
[\bar \ep] = \frac{{\rm length}^2}{{\rm time}^3}
$$

Eq.(\ref{magic}) may be used to estimate orders of magnitude of
basic quantities, e.g. the typical turnover time,
in the inertial range. The rules are that
these quantities depend on the scale $l$ only via a naive
dimensional analysis using eq.(\ref{magic}). Hence, at scale $l$, 
the typical velocity is $u_l\simeq \bar \ep^{1/3}\, l^{1/3}$
and the typical time is $\tau_l\simeq \bar \ep^{-1/3}\, l^{2/3}$.
The typical velocity gradient is 
$$
(\nabla u)_l\simeq u_l/l\simeq \bar \ep^{1/3}\, l^{-2/3}
$$
Note that it diverges as $l\to 0$ supporting the idea that
fully developed turbulence is supported by weak solutions,
those with possible singularities of the Navier-Stokes equation.
This singular behavior is only expected in
the inertial range, or equivalently only in the inviscid limit
of infinite Reynold's number.

The dissipative scale $\eta$ may be estimated by looking at the 
scale at which the advection $(u\cdot\nabla)u$ becomes of the
order of the dissipation $-\nu \nabla^2u$. This
gives $\nu u_\eta/\eta^2\simeq u^2_\eta/\eta$ and
$$
\eta \simeq \nu^{3/4}\, \bar \ep^{-1/4}
$$
As expected it vanishes as $\nu$ goes to zero, illustrating that
$\eta$, or $\nu$, is an ultraviolet cut-off. Below this cut-off
the advection may be neglected in the Navier-Stokes
equation and the velocity field becomes more regular.
Note that the energy dissipated by unit of time, which
may be evaluated as $-\nu (\nabla u)^2$ taken at the scale $\eta$, is:
$$ 
\nu\, (\nabla u)^2_\eta \simeq \bar \ep
$$
So that the energy dissipation rate is equal to the energy transfer rate,
as expected by energy conservation.

The Reynold's number at scale $l$ is $\CR_l = u_l\,l/\nu$
so that $\CR_l\simeq \bar \ep^{1/3}\, l^{4/3}/\nu$.
The system Reynold's number may be taken as $\CR_l$ at the
large integral scale $L$, $\CR e = \CR_L$, whereas the dissipation
scale is such that $\CR_\eta\simeq 1$. The inertial domain 
scale with the Reynold's number as:
$$
\frac{L}{\eta} \simeq \CR e ^{3/4}
$$
As it should be, the inertial range increases with the Reynold's number.

\subsection{Modeling statistical turbulence.}
Random forcing and stationary measure. \\
Energy balance and mean dissipation rate.\\
More on inertial range and Kolmogorov theory.\\

Universality in turbulence is expected to occur in the
inertial range only statistically,
in the sense that statistics of turbulent data extracted from
repeated experiments should be more or less independent on
the precise experimental set-up, eg on the ways the energy
is injected in the turbulent baths. 
To theoretically model these repeated experiments one 
considers random initial data and
random forcing in the Navier-Stokes equation (\ref{nsequa}).
By universality the inertial range statistics should then
be independent on the precise statistics chosen for the
force. The simplest is to choose the force to be
gaussian with zero mean and covariance:
\debut
\vev{f^j(x,t)\, f^k(y,s)} = C_L^{jk}(x-y)\,\de(t-s)
\label{force}
\fin
with $\nabla_j C^{jk}_L(x)=0$ to ensure transversality of the force. 
To mimic the fact that the energy is injected
at a large scale $L$, the covariance $C^{jk}_L(x)$, which can
be chosen in the form $C^{jk}(x/L)$, varies on scale $L$.
It is approximately constant up to scale $L$ and decreases
exponentially beyond $L$. One may choose $C^{jk}(x)$ to be a gaussian function.

The velocity field then becomes a random field and
one is interested in its multipoint correlations,
and more generally, in its various probability distribution functions.
The latter may be defined as:
$$
\vev{\de(u(x_1,t)-v_1)\cdots \de(u(x_n,t)-v_n)}
$$
These are functions of the positions and times at which
the velocity is measured. However, one expects that
at sufficient large time the turbulent systems 
reach a steady state, independent of the initial data.
Statistical stationarity means that the equal time
correlation functions are time independent:
\debut
\frac{d}{dt} \vev{u(x_1,t)\cdots u(x_n,t) } =0 \non
\fin

Stationarity implies a balance between the energy injected
into the system and the energy dissipated by viscous processes.
Indeed averaging the energy balance eq.(\ref{bilan}) with
$\frac{d}{dt}\vev{u^2/2}=0$ yields:
\debut
\vev{f\cdot u} = \nu \vev{ (\nabla u)^2} \equiv \bar \ep
\label{balance}
\fin
The mean injection rate of energy $\vev{f\cdot u}$ should
be equal to the mean dissipation rate $\nu \vev{ (\nabla u)^2}$.
As in previous section the energy is injected at large
scales of order $L$ and it is expected to be
dissipated at small scales of order $\eta$: the energy transfer
through the scale being without loss. 

The inertial range and energy cascade may then be formulated in a more 
precise way by introducing $\bar \ep_{\leq K}$, the energy dissipated
by modes with wave numbers of modulus less than $K$, and $w_{\leq K}$,
the energy injected into modes of wave numbers less than $K$.
They are defined by:
\debut
\bar \ep_{\leq K} &=& \int_{|k|\leq K}\frac{d^3k}{(2\pi)^3}\ 
\int d^3x\, \nu \vev{\nabla u(x)\cdot \nabla u(0)}e^{-ik\cdot x} \non\\
\bar w_{\leq K} &=& \int_{|k|\leq K}\frac{d^3k}{(2\pi)^3}\ 
\int d^3x\, \vev{u(x)\cdot f(0)}e^{-ik\cdot x}\non
\fin
The fact that the energy is injected at large scale means  
that $w_{\leq K}$ is approximately constant and equal
to $\bar \ep$ as soon as $K\geq 1/L$. The fact that the energy
is only dissipated at small scale means that $\bar \ep_K$ 
approximately vanishes at scales $K$ less than $1/\eta$ and 
becomes of order $\bar \ep$ for $K\geq 1/\eta$.
The flux of energy through the surface of wave numbers of
modulus $K$ is the difference:
\debut
\bar \pi_{\leq K} = \bar w_{\leq K}-\bar \ep_{\leq K} \non
\fin
The cascade picture then means that $\bar \pi_{\leq K}$ is approximately
constant for $K$ in the inertial range, 
$1/L \ll K \ll 1/\eta$. Looking for constant energy flux 
is an experimental signal for fully developed turbulence. 

The injected energy $\bar w_{\leq K}$ may be linked to the Fourier
transform of the force covariance (\ref{force}). One has:
\debut
\bar w_{\leq K} = \half \int_{|k|\leq K}\frac{d^3k}{(2\pi)^3}\ 
\int d^3x\, e^{-ik\cdot x}\, {\rm tr}C_L(x) \non
\fin
So choosing the energy injected spectrum as described above specifies 
the choice of the force covariance.

\vskip 1.0 truecm
$$\epsfbox{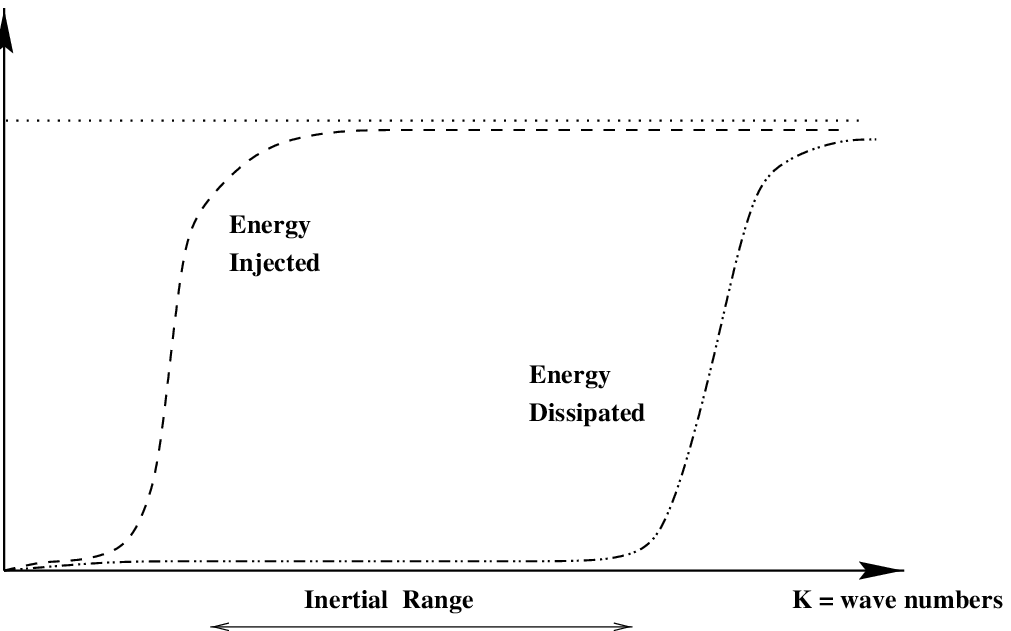} $$

\centerline{Figure 1: Detailed energy balance.}

\vskip 1.0 truecm

Velocity correlations in the inertial range $\eta \ll |x| \ll L$
are extracted by looking at the short distance limit 
in the inviscid correlators. Namely, to satisfy $\eta \ll |x|$
one first takes the inviscid limit $\nu \to 0$ at fixed positions $x$ 
and fixed integral scale $L$, and then, one takes the short 
distance limit at fixed $L$. Assuming the existence of scaling laws
for the $n$-point correlation functions means assuming
the existence of exponents $\xi_n$ such that the following
limits exist:
\debut
\lim_{\la\to 0} \lim_{\nu\to 0}\ \la^{-\xi_n} 
\vev{u(\la\,x_1,t)\cdots u(\la\,x_n,t) } \label{inviscid}
\fin
Remark that since in the inviscid limit 
there is no other scale than $L$ and the positions, taking the last
short distance limit is equivalent 
to sending the infrared cut-off to infinity.
Inverting the limit order would not give the same answer as it would
describe correlations in the dissipative domain.
Since $\nu$ plays a role of ultraviolet cut-off,
velocity correlations behave very differently at finite viscosity
and in the inviscid limit:
they are smooth at short distance at $\nu$ finite
but their derivatives develop ultraviolet singularities 
in the inviscid limit.

In his seminal 1941 paper \cite{K41}, Kolmogorov made a crucial step
by postulating the universal character of the velocity
statistics in the inertial range. In a simplifying way,
{\it Kolmogorov scaling theory}, known as K41, 
but not to be confused with K-theory, 
is based on the assumptions that
there is a constant energy cascade through some inertial range
and that the inertial range velocity correlators
are universal in the sense that they do not depend
on how the energy is injected, ie. on the force,
but only on the positions at which the velocities are taken and
on the mean energy transfer rate $\bar \ep$.
As a consequence, the structure functions 
$S_n(x)=\vev{ [(u(x)-u(0))\cdot \hat x]^n}$ are completely
determined by dimensional analysis up to constants:
\debut
S_n(x)\equiv\vev{ [(u(x)-u(0))\cdot \hat x]^n}= C_n\ (\bar \ep\, |x|)^{n/3}
\label{structure}
\fin
Compare this with eq.(\ref{magic}).
Universality implies that the constants $C_n$ are independent of the forcing.

The K41 theory also gives  a prediction for
the energy spectrum $E(k)$,
\debut
E(k) \propto \bar \ep^{2/3}\, k^{-5/3} 
\label{spectre}
\fin
with $E(k)$ the energy in modes of wave numbers of modulus $k$,
$$
E(k)\, dk = \int_{|k|=K}\frac{d^3k}{(2\pi)^3}\ 
\int d^3x\, \vev{u(x)\cdot u(0)}e^{-ik\cdot x}
$$

Eq.(\ref{structure}) is exact for $n=3$, see eq.(\ref{exact3}) below, 
but the scaling it implies for the higher order structure functions 
seems to be invalidated by real and numerical experiments.

\subsection{Correlation functions and scalings.}
Exact three point function.
\\
Anomalous scaling, universality and the large $L$ limit.\\

The simplest exact result on turbulence is a formula
for the order three structure function which supports
Kolmogorov scaling. At it should be clear from
previous sections it is a consequence of constant energy 
transfer. It follows from the Navier-Stokes
by imposing the stationarity of the velocity two-point
function: $\d_t \vev{u(x)\cdot u(y)}=0$.
This is an exercise worth doing in details.

Consider the Navier-Stokes equation with random
forcing with covariance (\ref{force}). It implies that
\debut
u(x,t+\de t)-u(x,t) &=&\int_t^{t+\de t} \d_s u(x,s)ds \non\\
&=& \[-(u\cdot \nabla)u+\nu \nabla^2u-\nabla p\](x,t)\de t
+ \int_t^{t+\de t} f(x,s)ds + \CO(\de t^2)   \non
\fin
Since the force is delta correlated in time, the term
linear in $f$ is of order $\de t^{1/2}$. Inserting this formula
into $\vev{u(x,t+\de t)\cdot u(y,t+\de t)}$ and keeping
term of order $\de t$ only gives:
\debut
\d_t\vev{u(x,t)\cdot u(y,t)} &=&
\vev{\[-(u\cdot \nabla)u-\nabla p\](x,t)\cdot u(y,t)}
+ \nu \vev{\nabla^2u(x,t)\cdot u(y,t)} +(x\leftrightarrow y) \non\\
&&+ {\rm tr}C_L(x-y) \non
\fin
The term proportional to $\vev{f(x,s)\cdot u(y,t)}$ vanishes
since by causality $u(y,t)$ is independent of $f(x,s)$ for $s>t$.
The term with the pressure vanishes by incompressibility 
and others may be simplified using translation invariance:
\debut
\vev{\nabla p(x,t)\cdot u(y,t)}&=&-\vev{p(x,t)\nabla \cdot u(y,t)}=0\non\\
\nu \vev{\nabla^2u(x,t)\cdot u(y,t)}&=&
-\nu \vev{\nabla u(x,t)\cdot\nabla  u(y,t)}\non\\
\vev{(u\cdot \nabla)u(x,t)\cdot u(y,t)}+(x\leftrightarrow y)
&=&-\inv{2} \nabla_x\cdot \vev{(u(x,t)-u(y,t)\, ((u(x,t)-u(y,t))^2}\non
\fin
Hence stationarity of the equal time two point function gives
(we drop the time label to simplify the notation):
\debut
-\inv{4} \nabla_x\cdot \vev{(u(x)-u(y)\, ((u(x)-u(y))^2} 
+\nu \vev{\nabla u(x)\cdot\nabla  u(y)}
=\half {\rm tr}C_L(x-y) \label{2stat}
\fin
We shall now take the limits $\nu\to 0$ and $x\to y$ in two different
orders. Take first the coincident point limit at finite viscosity.
The first term in eq.(\ref{2stat}) then vanishes since velocity
correlations are smooth at $\nu$ finite. Sending $\nu$ to zero afterwards
we get:
\debut
\lim_{\nu\to 0}\lim_{x\to y}\, \nu \vev{\nabla u(x)\cdot\nabla  u(y)}
=\half {\rm tr}C_L(0) \equiv \bar \ep \label{dissip1}
\fin
Taking the limit in the reverse order, first $\nu\to 0$ and then
$y \to x$, the second term in eq.(\ref{2stat}) vanishes and we get:
\debut
-\inv{4}\lim_{x\to y}\lim_{\nu\to 0}\,
\nabla_x\cdot \vev{(u(x)-u(y)\, ((u(x)-u(y))^2}=\bar \ep
\label{dissip3}
\fin
Assuming isotropy this implies that at short distance 
\debut
\vev{(u^i(x)-u^i(0))(u^j(x)-u^j(0))(u^k(x)-u^k(0))}
=-\frac{4}{15}\, \bar \ep\, \(\de^{ij} r^k + \de^{jk} r^i+ \de^{ki} r^j\)
\label{dissip2}
\fin
with $r=|x|$. Equivalently,
\debut
S_3(x)=\vev{ [(u(x)-u(0))\cdot \hat x]^3}= -\frac{4}{5}\ \bar \ep\, r
\label{exact3}
\fin
This is known as {\it Kolmogorov four-fifths law}. 
As it is clear from eq.(\ref{dissip1}), $\bar \ep$ is the mean dissipation rate.
Remark that the Kolmogorov four-fifths law essentially follows
from the cinematics underlying the Navier-Stokes equation.

Of course eq.(\ref{exact3}) matches Kolmogorov scaling (\ref{structure})
with $C_3=-4/5$. In dimension $d$, it would be $-\frac{12}{d(d+2)}$.
However, there are strong experimental as well as numerical evidences 
that other structure functions do not follow Kolmogorov scaling but
$$
S_n(x) \propto r^{\xi_n}
$$
with $\xi_n\not= n/3$. Sample of available data are
$\xi_4\simeq 1.28$, $\xi_5\simeq 1.53$, $\xi_6\simeq 1.77$, etc.
$\xi_{2n}$, $2n\geq 3$, should be less than $2n/3$. 
See e.g. \cite{frisch} for a discussion of the experimental data.

In the inviscid limit, the only dimensionfull parameter on which
the inviscid correlations may depend are the positions $x$, the
integral scale length $L$ and the mean dissipation rate $\bar \ep$.
These correlations may also a priori depend on the shape
of the force covariance via $\bar \ep^{-1} C^{jk}(x/L)$
but this is a dimensionless parameter. So on dimensional ground,
the $n$-point inviscid correlations are of
the form $$(\bar \ep\, |x|)^{n/3}\, F_n(x/L).$$ 
Kolmogorov hypothesis may then be seen as assuming that
the large $L\to \infty$ limit exists whereas breakdown of
Kolmogorov scaling on contrary means that $F_n(x/L)$ diverge
for large $L$. In such case the structure functions 
behave as:
\debut
S_n(x) = A_n\, (\bar \ep\, |x|)^{n/3}\, (|x|/L)^{\xi_n-n/3}
\quad {\rm as} \quad L\to\infty  \label{deviation}
\fin
Recall that $\xi_n<n/3$ so $S_n(x)$ grows with $L$.
Universality is expected to be restored in the sense that
the anomalous scaling dimensions $\xi_n$ are independent
of the forcing but the dimensionless amplitudes $A_n$ depend on the
details of the forcing via its dimensionless covariance
and are therefore not universal.
Since $\xi_n<n/3$ the higher moments (\ref{deviation}) of the velocity 
differences become larger and larger than their gaussian
values as one goes to shorter and shorter scales. This means that
the probability to have large velocity differences 
increase at short scales. This phenomena is called intermittency.

\vfill \eject

\section{Deeper in turbulence flows.}

\subsection{Turbulence in 2d.}
Inverse versus direct cascades. \\
Enstrophy dissipation but no energy dissipation.\\
Kraichnan's scaling, KK-theory and exact $3$-point function.\\

A special feature which distinguishes two dimensional
from three dimensional fluid mechanics is 
the conservation of vorticity moments in the inviscid limit. 
The vorticity $\om$ is defined by $\om=\ep_{ij}\d_iu_j$ with
$\ep_{ij}=-\ep_{ji}$, $\ep_{12}=1$. It is transported by the fluid
since it satisfies
\debut
\d_t\om +(u\cdot\nabla)\om - \nu \nabla^2 \om = F \label{omeq}
\fin
where $F=\ep_{ij}\d_if_j$ with $f_j$ the force applied to the
Navier-Stokes equation (\ref{nsequa}). As a consequence, 
$\d_t \int \om^n=0$ in absence of force and at zero viscosity. 
The second moment is called the enstrophy so that the enstrophy
density is $\Om=\half \om^2$.
The  2d inviscid Navier-Stokes equation admits
two quadratic conserved quantities: the energy $\int \frac{u^2}{2}$
and the enstrophy $\int \frac{\om^2}{2}$.

As first pointed out by Kraichnan in a remarkable paper \cite{kraich},
this opens the possibility for quite different scenario 
for the behavior of turbulent flows in two and three dimensions, 
see also \cite{batch}.
As argued by Kraichnan, if energy and enstrophy
density are injected at a scale $L$, with respective rate $\bar \ep$ and 
$\bar \eta_w\simeq \bar \ep L^{-2}$, the turbulent system should
react such that the energy flows toward the large scales and the enstrophy
towards the small scales. As this energy flow is quite the opposite
to the one involved in Kolmogorov's picture for 3d turbulence, 
one usually refers to the infrared energy flow as {\it the inverse 
cascade} and to the ultraviolet enstrophy flow as {\it the direct cascade}. 

The fact that energy has to escape to the large scales may be
understood by looking at the time variation
of the energy  and enstrophy in absence of forcing,
\debut
\d_t \int \frac{u^2}{2} &=& -\nu \int (\nabla u)^2
= -\nu \int \,\om^2 \non\\
\d_t \int \frac{\om^2}{2} &=& -\nu \int (\nabla \om)^2
\non
\fin
In three dimensions, the dissipation rate $\d_tu^2/2$ is finite in 
the inviscid limit so that the limit $\nu\to 0$ is accompanied by an increase
of the mean square vorticity. This cannot happen in two dimensions
because the enstrophy is bounded by its initial value since it can only decrease. 
The total energy is then constant in the limit $\nu\to 0$,
in particular it is not dissipated at small scales. 
Furtheremore, any transfer of energy as to be compensated
by a large transfer of energy towards the large scale (small momentum $k$)
in order for the enstrophy to decrease, $\int dk \,\d_tE(k)\,k^2<0$,
and the energy to be conserved $\int dk \,\d_tE(k)=0$.
At the same time, the enstrophy will be transfered towards the
small scale (large $k$) at which it will be dissipated, 
since the dissipative term $-\nu\nabla^2 \om$ dominates over
the advection term $u\cdot\nabla \om$ at small scale.

The energy cascade towards large scales is the inverse cascade.
It is characterized by the mean energy transfer $\bar \ep$.
Scaling arguments lead to Kolmogorov's spectrum,
with $E(k)\sim \bar \ep^{2/3}\,k^{-5/3}$ for the energy and 
$(\de u)(r) \sim (\bar \ep r)^{1/3}$
for the variation of the velocity on scale $r$.

The enstrophy cascade towards small scales is the direct cascade.
It is characterized by the mean enstrophy  transfer $\bar \eta_w$.
Since $\bar \eta_w$ has dimension 
$$
[\bar \eta_w] = {\rm time}^{-3},
$$
scaling arguments give the so-called Kraichnan's spectrum with 
\debut
E(k)\simeq \bar \eta_w^{2/3}\, k^{-3} \label{krachE}
\fin
for the energy and 
\debut
(\de u)(r) \simeq \bar \eta_w^{1/3}\, r  \label{krachU}
\fin 
for the velocity variation.
In particular, the velocity field is expected to be much smoother
in 2d turbulence than in 3d turbulence.
The scaling behaviors (\ref{krachE}) and (\ref{krachU}) 
cannot be true at the same time since the scaling (\ref{krachE}) for the
energy implies logarithmic behavior for the velocity, 
$(\de u)^2(r) \simeq \bar \eta_w^{2/3}\, r^2\log r$, whereas
the scaling (\ref{krachU}) for the velocity implies a more
regular energy spectrum, $E(k)\simeq k^{-3-\al}$ with $\al>0$.
One sometimes refers to eqs.(\ref{krachE},\ref{krachU}) to
Kolmogorov-Kraichnan's theory, but not to be confused
with KK-theory.

As in 3d one may derive exact result for the three-point structure functions
under the hypothesis that there is no inviscid energy dissipation,
$$
\lim_{\nu\to 0} 
\nu \vev{ (\nabla u)^2 } = 0 
$$
but there is enstrophy dissipation,
$$
\lim_{\nu \to 0} \nu \vev{ (\nabla \om)^2 }=\bar \eta_w
$$
As a consequence, not all correlation functions are expected to
reach a stationary regime but only the galilean invariant ones.

The derivation of this exact result is similar to the derivation
of Kolmogorov's $4/5$~law, so we shall be more sketchy.
The two point velocity correlations $\vev{u(x)\cdot u(0) }$
satisfies the same equation of motion as in 3d. 
Hence, in absence of energy dissipation  in the inviscid limit,
the mean energy increases linearly with time according to:
\debut
\d_t \vev{u^2/2}=  \bar \ep 
\label{energy}
\fin
So, $\vev{\frac{u^2}{2}}=  \bar \ep\, t$ up to a constant,
and $\bar \ep=\half\,{\rm tr}\, C_L(0)$ is indeed the energy injection rate.
This is simply the obvious statement
that in absence of energy dissipation, and/or in absence of
friction or other processes by which the energy may escape,
all energy injected into the system is transfered to the fluid.
It is expected to be transfered to the mode with the smallest possible
momentum, the so-called condensate \cite{kraich}. 
In particular eq.(\ref{energy}) shows that in absence of energy dissipation
a stationary state cannot be reached although structure
functions may converge at large time.
This is one important difference between 2d and 3d turbulence.

Assuming that galilean invariant correlations, say  $\vev{ (\de u)^2(x)}$
or $\vev{\om(x)\om(0)}$, are  stationary, one gets two inviscid equations:
\debut
\half\nabla_x^k \vev{ (\de u^k)(x)\ (\de u)^2(x)} 
= 2\bar \ep - {\rm tr}\, C_L(x)
\label{3point0}
\fin
and
\debut
-\half \nabla_x^k \vev{ (\de u^k)(x)\ (\de \om)^2(x)}
=G_L(x)  \non
\fin
with  $G_L=-\nabla^2 {\rm tr}\,C_L$. One has $G_L(0)= 2\bar \eta_w$.
These equations are enough to determine the three point structure functions.
Eqs.(\ref{3point0}) slightly differs from its 3d analogue (\ref{2stat})
by the extra term proportional to $\bar \ep$ in its r.h.s. which
takes into account for the energy increase.

In the direct enstrophy cascade, $x\to 0$, this gives \cite{db99}:
\debut
\vev{ (\de u)_{\|}^3 } = \vev{ (\de u)_{\|} (\de u)_{\bot}^2 }
\simeq +\frac{{\bar \eta_w}}{8}\ r^3
\label{law}
\fin
for the transverse and longitudinal correlations.
$\bar \eta_w$ is equal to the mean enstrophy dissipation rate.
Thus, as expected the 3-point velocity functions, which only depend
on the enstrophy injection rate, are smooth and universal in the direct
cascade. Eq.(\ref{law}) may be called the "$+1/8$ law".

In the inverse energy  cascade, $x\to \infty$, one gets:
\debut
\vev{ (\de u)_{\|}^3 } =3\vev{ (\de u)_{\|} (\de u)_{\bot}^2 }
\simeq +\frac{3\bar \ep}{2}\ r 
\label{lawbis}
\fin
with $\bar \ep$ the mean energy injection rate.
Of course this gives Kolmogorov's $+3/2$ law for the longitudinal
statistics.

Although there are necessarely some deviations from  Kolmogorov's `mean field' 
theory either in the energy spectrum or in the velocity statitics 
in the direct cascade, they seem  to be less pronounced in 2d than in 3d.
A proposition for logarithmic deviations in the direct cascade was 
presented in \cite{FaLe}.
Two dimensional turbulence has recently been observed in remarkable experiments \cite{Tab}.
These experimental data seem to indicate almost an absence of, or at least very weak,
deviations from Kolomogorov-Kraichnan scaling both in the direct and in the inverse
cascade. They also show that the probability distribution functions of
the even order structure functions are close to gaussian distributions.
More experimental as well as numerical recent results on two dimensional cascades 
are avalaible in refs.\cite{also2d}. These give support for very tiny, if any,
deviations from Kolmogorov and Kraichnan energy spectrum in the inverse and
direct cascade respectively. But they also indicate that correlation
functions don't exactly obey gaussian distributions.
These experiments were done in cells, the bottom of which induce friction
on the turbulent fluids. This friction is necessary for reaching a stationary regime
since there is no dissipative anomaly in two-dimension. The influence of friction
on the direct cascade has been analysed numerically in \cite{nurfric} 
and  different scenario have been discussed in \cite{fric}.
An attempt to use conformal field theory for constructing zero modes in
2d turbulence was described in \cite{Pol}.

\subsection{Dissipation and dissipative anomalies.}
Stationarity equation again and dissipative anomaly.\\
Dissipation for weak solutions.\\

As it has been apparent in previous section, fully
developed turbulence is supported by velocity fields
which are non-smooth in the inviscid limit.
This may be argued either from the scaling argument (\ref{magic})
or using the energy balance (\ref{balance}). 
At finite viscosity velocity correlations are smooth
because $\nu$ and the dissipation term in the Navier-Stokes
act as ultraviolet regulators.
But as $\nu\to 0$ correlations develop singularities. 
These are not present in the velocity correlations, which
are still finite in the coincident point limit (because we
may measure velocity moments), but they are present in correlations
of derivatives of the velocity. This a consequence of the fact that
velocity differences $\de u(x)$ between two neighbour points
separated by a distance $x$ scale as  $\de u(x)\sim x^{1/3}$
in Kolmogorov approximation. This is also a consequence
of the stationarity condition which imposes the dissipation to be non zero,
$$
\lim_{\nu \to 0} \nu \vev{(\nabla u)^2} = \bar \ep
$$
showing that gradients of velocities are diverging at coincident points
in the inviscid limit.  
More generally, the dissipation field defined as
\debut
\ep^{ij}(x) = \lim_{\nu\to 0}\,\nu\, (\nabla^ku^i)(x)(\nabla^ku^j)(x) \label{dissop}
\fin
is non vanishing in the inviscid correlators. 
This fact is called {\it the dissipative anomaly}.

By manipulating the Navier-Stokes equation as in previous section, 
see eqs.(\ref{dissip1},\ref{dissip3}), one may derive an alternative expression
for the dissipation field directly as an operator in the inviscid
theory:
\debut
{\rm tr} \ep(x) =
-\inv{4}\lim_{x\to y}\lim_{\nu\to 0}\,
\nabla_x\cdot \vev{(u(x)-u(y)\, ((u(x)-u(y))^2}
\label{opediss}
\fin
This relation may be thought as some part of the short 
distance operator product expansion of velocity fields in the inviscid limit.
Again it illustrates the fact that fully developed turbulence is supported
by non-smooth alias weak solutions of the Navier-Stokes equation.

To illustrate the occurrence of the dissipative anomaly 
let us write the stationarity equation for arbitrary 
correlation functions $\vev{ \prod_n\, F_n[u(x_n)]}$
with $F_n$ any functions of the velocity $u(x_n)$
without derivatives. It reads:
\debut
&&\sum_k\, \vev{\[\((u\cdot\nabla)u^{j_k} +\nabla^{j_k} p\)(x_k)
\frac{\de}{\de u_{j_k}(x_k)}
+\ep^{i_kj_k}(x_k)\frac{\de^2}{\de u_{i_k}(x_k)\de u_{j_k}(x_k)}\]\,
\,\prod_n\, F_n[u(x_n)] }\non\\
&& ~~~~~ =\half\sum_{k,l} C_L^{i_kj_l}(x_k-x_l)\, 
\vev{\frac{\de^2}{\de u_{i_k}(x_k)\de u_{j_l}(x_l)}\,
 \prod_n\, F_n[u(x_n)]}
\label{station2}
\fin
The particular case with $F[u]=u$ reproduces eq.(\ref{dissip3}).
The presence of the dissipative field in this equation
shows that the set of relations satisfied by the velocity
correlations as a consequence of the stationarity do not form
a close set of equations. 
Eq.(\ref{station2}) are some kind of equations of motion.
Note that the precise form of the r.h.s. follows from the
hypothesis that the force has a gaussian statistics. 

If there are universalities
of inertial range velocity correlations,
these correlations should not depend on the choice of the force
statistics, and thus should be characterized by the l.h.s.
of the equations of motion (\ref{station2}).
As we shall see this is the scenario we shall encounter 
in Kraichnan's model: the inertial statistics
will be dominated by {\it zero mode solutions} of the l.h.s. of 
the stationarity equation.

Let us do this exercise in details for the two-point
functions $\vev{ F_1[u(x_1)]\, F_2[u(x_2)]}$.
The generalization to an arbitrary number of points is simple.
It is described in a more abstract way in the following section.
The stationarity condition for the two-point function is:
\debut
\d_t \vev{ F_1[u(x_1)]\, F_2[u(x_2)]} = 0 \non
\fin
Using the Navier-Stokes equation and the chain rule, 
$\d_t F[u]=(\d_tu)\cdot \de F[u]$, it becomes:
\debut
&& ~~~~~~~ \vev{\[-(u\cdot\nabla)u^j -\nabla^j p\](x_1)
\, \de_jF_1[u(x_1)]\, F_2[u(x_2)]} \label{develop}\\
&+& \nu \vev{ \nabla^2u^j(x_1)\, \de_j F_1[u(x_1)]\, F[u(x_2)]}
+ \vev{f^j(x_1)\, \de_j F_1[u(x_1)]\, F[u(x_2)]} 
+(x_1 \leftrightarrow x_2) =0 \non
\fin
with $\de_{j_k} F[u]$ the derivative of $F$ with respect to $u^{j_k}(x_k)$.
The terms in the first line do not pose problems and are well defined 
in the inviscid limit. The second term, which would vanish naively
in the inviscid limit, is actually non zero due to the dissipative
anomaly. Using translation invariance to distribute one of the
gradient operator on the function $F$, it may be rewritten 
in the inviscid limit as:
\debut
&~&-\nu \vev{ \nabla^2u^j(x_1)\, \de_j F_1[u(x_1)]\, F[u(x_2)} \non\\
&=& \nu \vev{ (\nabla u^i(x_1)\cdot\nabla u^j(x_1))\, 
\de_i\de_j F_1[u(x_1)]\, F[u(x_2)} + \nu
\nabla_{x_1}^k\nabla_{x_2}^k \vev{ F_1[u(x_1)]\, F_2[u(x_2)]}\non\\
&=& \vev{ \ep^{ij}(x_1)\de_i\de_jF_1[u(x_1)]\,  F[u(x_2)} \non
\fin
with $\ep^{ij}(x)$ defined in eq.(\ref{dissop}). We used the fact that
the second term in the second line vanishes in the inviscid limit
because the correlation $\vev{ F_1[u(x_1)]\, F_2[u(x_2)]}$ and hence
its derivatives are finite at $\nu=0$. 
We now have to compute the term in (\ref{develop})
involving the force. Since the force is gaussian with 
covariance (\ref{force}), functionally integrating by part gives:
$$
\vev{f^i(x)\ G[u(y)]} = \int d^3z C_L^{ik}(x-z) \vev{\frac{\de G[u(y)]}{\de f^k(z)}}
$$
for any functional $G$ of $u$.
So we have to evaluate the functional derivative 
$\frac{\de u^j(y,t)}{\de f^k(z,s)}$
in the equal time limit $t=s$. By causality this derivative vanishes for
$s>t$. Derivating the Navier-Stokes equation shows that it satisfies a first
order differential equation
$$
\[\d_t + \CN[u](y)\] \frac{\de u^j(y,t)}{\de f^k(z,s)}
= \de^{jk} \de(y-z)\de(t-s)
$$
for some differential operator $\CN[u](y)$ depending on $u$. 
As a consequence.
$$
\frac{\de u^j(y,t)]}{\de f^k(z,s)}=\th(t-s)\, G^{jk}(z,y|t,s)
$$
with $\th(t)$ the step function and $G$ the solution of the linear equation
$\[\d_t + \CN[u](y)\]G^{jk}(z,y|t,s)=0$ with the initial condition
$G^{jk}(z,y|t,s)_{t=s}=\de^{jk}\de(z-y)$. Taking the equal time limit,
with $\th(0)=\half$, we get:
\debut
\frac{\de u^j(y,t)]}{\de f^k(z,s)}\Big\vert_{t=s}=\half\de^{jk}(z-y)
\label{usurf}
\fin
This gives,
\debut
\vev{f^i(x)\ G[u(y)]} =\half
C_L^{ik}(x-y) \, \vev{\de_k G[u(y)]}
\label{fGdeu}
\fin
Choosing $\th(0)=\half$ is a little bit arbitrary but any other choice will
give the same result in eq.(\ref{develop}), assuming translation invariance.
Gathering everything into eq.(\ref{develop}) gives the
stationarity equation (\ref{station2}).

\subsection{Fokker-Planck equations.}
Fokker-Planck equation.\\
MSR formalism.\\

The Navier-Stokes equation with a random forcing belongs to the class of
stochastic equations of the form
\debut
\d_t q = V(q) + f   \label{stocha}
\fin
with $V(q)$ the forces applied to the dynamical variables $q$ and $f$
gaussian noices with covariance, $\vev{f(t)\,f(s)}=
\Ga\, \de(t-s)$, with $\Ga$ some symmetric matrix. 
For the Navier-Stokes equation the
variables $q$  are the velocity fields $u(x,t)$.
The above method to derive stationarity equations
applies as well to these stochastic equations.
Time derivative of equal time correlation functions are:
$$\d_t\vev{\prod_j q_j}=
\sum_j \vev{V(q_j)\,\prod_{k\not= j} q_k}
+ \sum_j \vev{f_j\,\prod_{k\not= j} q_k }
$$
The last term may be integrated by part since the noices are gaussian:
$$
\vev{f^j\,\prod_{k\not= j} q_k }= \Ga^{jl} 
\vev{\frac{\de }{\de f^l}\prod_{k\not= j} q_k }
$$  
Furthermore, since eq.(\ref{stocha}) is a first order causal differential equation
on still has that the variation $\de q(t)$ is independent of $f(s)$ for
$s>t$ and that it satisfies the linear equation: $[\d_t+\de V(q)]\de q=\de f$. 
Thus $\frac{\de q_i(t)}{\de f_j(s)}\vert_{t=s}=\half\, \de_{ij}$,
with the convention that $\th(0)=\half$ as in eq.(\ref{usurf}). 
Hence, one gets :
\debut
\d_t\vev{\prod_j q_j}=
\sum_j \vev{V(q_j)\,\prod_{k\not= j} q_k}
+\half \sum_{i,j} \Ga_{ij}\, \vev{\prod_{k\not=i,j}q_k} \label{statstocha}
\fin
These are generalized Fokker-Planck equations.
Similarly, for any function $F(q)$,
$$
\d_t \vev{F(q)} = \vev{ V^j(q) \frac{\de F(q)}{\de q^j}}
+\half \Ga^{jl}\vev{ \frac{\de^2 F(q)}{\de q^j\de q^l}}
$$

When the force derives from an action such that
$ V(q) = -\Ga\, \frac{\de S}{\de q}$ for some action $S(q)$, 
a stationary measure  for eq.(\ref{stocha})
is provided by the Gibbs measure $dq\, \exp[-S(q)]$.

This does not apply to the turbulence problem because the force
does not derive from an action. The invariant measure is, of course,
much more difficult to find.

There however exists a path integral formalism to compute 
correlation functions. It is known as the MSR formalism \cite{msr}.
It consists in introducing in the path integral constraints 
imposing the equations of motion. Writing the stochastic equation
symbolically as in eq.(\ref{stocha}), it leads to the following
effective action for two sets of variables $\varphi$ and $q$:
\debut
S= i\int dt\, \varphi\cdot(\d_tq-V(q)) + 
\half \int dt\, (\varphi\cdot\Ga\cdot\varphi )
\label{mrs}
\fin
The variables $\varphi$ are introduced as Lagrange multipliers
imposing the equations of motion.
This applies to the Navier-Stokes equation with a 
random gaussian forcing. However it does not provide an efficient way
to determine the stationary measure, because there is no known efficient way
to deal with the non-linear advection term which is dominating
in the turbulent regime.

\subsection{Multifractal models.}
velocity $\de u_l(r)\simeq u_0 (l/l_0)^h$ on domain of dimension $D(h)$.\\
anomalous exponents.\\
examples: $\beta$ and bifractal models, saturation $\xi_\infty$.\\

Intermittency  means that the probability for having large velocity
fluctuations increases at short scale. 
Multifractal models are phenomenological models to encode the
fact that only part of the fluid participates to the
cascade: the amount of the participating fluid modes
decreasing with the scale \cite{fractal}.
Although these models do not explain the origin of intermittency
phenomena they present a simple way to link anomalous
scalings to geometrical properties of the  fluid behavior.

Let us first consider the simplest of the multifractal models
called the $\beta$-model.
In the spirit of Richardson cascade, let us suppose
that there is a hierarchy of eddies of scales
$l_n= l_0\varsigma^n$, $n=0,1,\cdots$, with $\varsigma<1$.
Now suppose that, contrary to the usual Richardson cascade, 
the volume $V_{l_n}$ occupied by the eddies decrease from one generation 
to the next by a factor $\beta$ so that
$$
V_{l_n} \simeq l_0^3\, \beta^n = l_0^3 \(l_n/l_0\)^{3-D}
$$
with $3-D = \log\beta/\log\varsigma>0$.
Demanding a constant energy flux $\bar \ep =\pi_l\simeq V_l\, u_l^3/ l$
yields
$$
\de u_l \simeq u_0\, \(l/l_0\)^{(D-2)/3}
$$
for the typical variation velocity $\de u_l$ at scale $l$, 
with $u_0^3=\bar \ep\, l_0$. Assuming that the probability to
find eddies of size $l$ at some point in the fluid 
is proportional to the volume occupied by them, the
moments of the velocity variation may be estimated as
$$
\vev{ (\de u_l)^p } \simeq u_0^p \, \(l/l_0\)^{\xi_p}
\quad {\rm with}\quad \xi_p =p/3 + (3-D)(1-p/3)
$$
which describe some deviations from Kolmogorov's scaling.

More generally, assume that for $x$ in a subset $\CV_h$ 
with ${\rm dim}\CV_h= D_h$ the velocity variation 
$\de u_l(x) = u(x+l)-u(x)$ scales for $l\to 0$ as
\debut
\de u_l(x) \simeq u_0\,\(\frac{l}{l_0}\)^h
\quad {\rm for}\quad x\in \CV_h , 
\quad {\rm dim}\CV_h=D_h \label{fract}
\fin
The possible scaling dimensions $h$ are supposed to belong
to some interval $h_{\rm min}\leq h \leq h_{\rm max}$, and
$D_h$ form the spectrum of fractal dimensions.

To compute the moments of the velocity variation one then
assumes that the probability to find the scaling behavior (\ref{fract})
for the velocity variation $\de u_l(x)$ is proportional 
to probability for the segment between $x$ and $x+l$
to cross a point in $\CV_h$. This probability is proportional
to the volume of $\CV_h$ thickened along its transverse directions
on a depth of order $l$. 

\vskip 1.0 truecm

$$\epsfbox{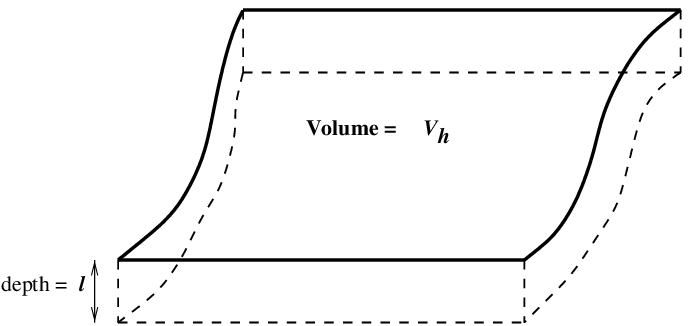} $$

\centerline{Figure 2 : Thickened fractal volume.}

\vskip 1.0 truecm

This volume scales as $\simeq l^{d-D_h}$
with $d$ the dimension of the ambient space, usually $d=3$.
Hence,
$$
\vev{(\de u_l)^p} \simeq u_0^p\,\int d\mu(h)\, (l/l_0)^{ph+d-D_h}
$$
with $d\mu(h)$ some measure encoding the probability
distribution of the fractal volumes $\CV_h$. In the short
distance limit, $l\to 0$, the above integral is dominated
by a saddle point and thus
\debut
\vev{(\de u_l)^p} \simeq u_0^p\,\(\frac{l}{l_0}\)^{\xi_p}
\quad {\rm with}\quad 
\xi_p= \min_h\( ph + d - D_h \) 
\label{fractal}
\fin
Since intermittency means that velocity variations become larger at
short scales, one expects that the spectrum of $h$ is such that
$h\leq h_{\rm max}=1/3$. Of course $0\leq D_h\leq d$ so that
$\xi_p\leq p/3$ as expected. 

The relation between the fractal dimensions $D_h$ and the
exponents $\xi_p$ is a Legendre transformation. It can be thus 
inverted as 
\debut
D_h = \min_p\( ph + d -\xi_p\) \label{fracinv}	
\fin
As it is clear, this geometrical construction does not
predict the spectrum of anomalous dimensions.

The $\beta$-model is a fractal model with a single fractal
dimension $h=(D-2)/3$.
The next simplest multifractal models are bifractal models with
$h$ taking only two values $h_0$ and $h_1$.
This is a case we shall encounter in one dimensional Burgers turbulence
in which $h_0=0$ with $D(0)=0$ (ie. points) and 
$h_1=1$ with $D(1)=1$ (ie. intervals). The anomalous dimensions are
then $\xi_p = \min(1,p)$. It saturates to the value $\xi_p=1$ as soon
$p$ is bigger than one.

The fact that the anomalous dimensions $\xi_p$ saturate 
to some finite value $\xi_\infty$ when $p\to \infty$ as an
interesting geometrical interpretation. Indeed, 
assuming that $h$ belongs to some interval so that $D_h$ is bounded,
then finiteness of $\xi_\infty$ is compatible with eq.(\ref{fractal})
only if  $h_{\rm min}=0$. From eq.(\ref{fract}), this means that
there exists a subset of the fluid volume $\CV_0$ of
dimension $D_0=d-\xi_\infty$ on which the velocity is 
discontinuous, since $\de u(x) \simeq  \CO(1)$ for $x\in \CV_0$.
In other words, finiteness of $\xi_\infty$ implies velocity shocks.

Multifractal models may also be used to estimate behaviors of other
quantities such as the dissipation. The inviscid relation
(\ref{opediss}) between the velocity and the dissipation field,
$\ep(x) \propto \d_l(\de_l u(x))^3$,  tells us that for
$x$ on a subset $\CV_h$ of dimension $D_h$ the dissipation
scales as:
$$
\de_l\, \ep(x) \simeq \bar \ep\, \(l/l_0\)^{3h-1}
\quad {\rm for}\quad x\in \CV_h
$$
Here, $\de_l\, \ep(x)$ may be defined by averaging $\ep(x)$ on a ball
of radius $l$ around $x$.
As it should $\de_l\, \ep(x)$ is singular as $l\to 0$, for $h_{\rm max}=1/3$.
Their moments behave as
\debut
\vev{\(\de_l\,\ep\)^p} \simeq {\bar \ep}^p\, (l/l_0)^{\eta_p}
\quad {\rm with}\quad
\eta_p=\min_h\( (3h-1)p+d-D_h\)= \xi_{3p}-p
\label{scadiss}
\fin 
Contrary to velocity correlations,  dissipation correlations 
decrease with the distances in the inertial range since $\xi_{3p}<p$.

\vfill \eject

\section{Scalar turbulence.}

\subsection{Transport  and Lagrangian trajectories.}
Transport equation.\\
Lagrangian flows.\\
Definition of the pdf's (backward/forward).\\

Scalar turbulence, which aims at describing transport phenomena
in turbulent flows \cite{obuk}, provides a simple toy model for
modeling turbulence statistically. 
The equation governing a passive scalar advected by a turbulent flow 
and subject to a small dissipation is:
\debut
\d_t T + \nabla\cdot(u\,T) -  \kappa \De T = f.
\label{eqT}
\fin
Here $T(x,t)$ represents the passive scalar, e.g. the density of another passive fluid.
The forcing term $f$ is here to compensate the energy dissipation caused
by the term proportional to the molecular diffusivity $\kappa$.
The velocity field $u$  with $\nabla\cdot u=0$ is supposed to be random.

The transport equation (\ref{eqT}) is linear and can thus be solved explicitly. 
Suppose for simplicity that at initial time,
$T(x,t_0)=T_0(x)$, the solution to eq.(\ref{eqT}) is then:
\debut
T(x,t)= \int_{t_0}^tdsdy\, R_\kappa(x,t|y,s)\, f(y,s)
+ \int dy\, R_\kappa(x,t|y,t_0)\, T_0(y)
\label{sol1}
\fin
with $R_\kappa$ the elementary solution of 
$(\d_t+\nabla\cdot u-\kappa\nabla^2)R_\kappa=0$
with initial condition, $R_\kappa(x,t_0|y,t_0)=\de(x-y)$. 

As a consequence of eq.(\ref{sol1}), correlation functions
of the scalar may formally be written as: 
\debut
\vev{\prod_{j=1}^N T(x_j,t)}
=\int^t_{-\infty} dsdy\, \vev{\prod_jR_\kappa(y_j,s_j)}\,\vev{\prod_jf(y_j,s_j)} 
\label{Tlagrange}
\fin
in the case with zero initial condition $T_0=0$.
The force and trajectory statistics have been factorized.

In the limit of zero molecular diffusivity, $\kappa\to 0$,
properties of the passive scalar are of course intimately related
to behaviors of passive particles advected by the fluid.
Consider a passive particle initially
at position $x_0$ at time $t_0$ in a turbulent fluid
and transported by it.
It follows a trajectory, called a Lagrangian trajectory,
whose equation of motion is the first order differential equation:
\debut
\dot x(t)\ = \ u(x(t),t)\quad {\rm with}\quad x(t_0)=x_0
\label{trajet}
\fin
with $u(x,t)$ the velocity field. When needed to specify the initial data,
we shall denote by $x(t|x_0,t_0)$ the solution of eq.(\ref{trajet}).

In the limit $\kappa\to 0$, the resolvent $R_\kappa$ may formally be expressed 
in terms of solutions of Lagrangian trajectories:
\debut
R_{\kappa=0}(x,t|y,s)=\de(x- x(t|y,s))	\label{defR}
\fin
with $x(t|y,s)$ the time $t$ position of the particle starting at $y$ at time $s$.
The solution (\ref{sol1}) then codes the fact at $\kappa=0$ the scalar is 
simply transported by the fluid. It provides an explicitly link between 
properties of the scalar and those of the Lagrangian trajectories.

Lagrangian trajectories may be described more precisely by introducing
their probability distributions. One may specify either the initial
or the final positions. In the former case, one looks for
the distribution of the final positions. This is encoded in
the forward probability distributions which are formally defined 
for $n$ particles as:
\debut
P_{\rm forward}^{(n)}(x,t|x_0,t_0)\, dx &\equiv& 
{\rm Proba}\(x(t)\in[x,x+dx];\, x(t_0)=x_0,\, t_0<t\) \non\\
&=&\vev{\prod_{j=1}^n \de(x_j- x(t_j|x_{0\,j},t_{0\, j}))}\, dx
\label{forward}
\fin
where $x(t|x_0,t_0)$ is the solution of the Lagrange equation (\ref{trajet})
with initial condition $x(t_0)=x_0$. This is the time $t$ position 
of the particle which was at $x_0$ at time $t_0$. 
The average is taken over the velocity field realizations. 

Conversely one may specify the final positions and look for
the distribution of the initial positions. These are formally 
defined as:
\debut
P_{\rm backward}^{(n)}(x,t|x_0,t_0)\, dx_0 &\equiv&
{\rm Proba}\(x(t_0)\in[x_0,x_0+dx_0];\, x(t)=x,\, t>t_0\) \non\\
&=& \vev{\prod_{j=1}^n \de(x_{0\,j}- \hat x(t_{0\,j}|x_j,t_j))}\, dx_0
\label{backward}
\fin
with $\hat x(t_0|x,t)$ the position of the trajectory at time $t_0$
which will be at $x$ at later time $t>t_0$.

Caution has to be taken when the velocity if not regular enough
since, as we shall discuss later, in  such cases Lagrangian trajectories 
may not be uniquely defined.
This is expected to be the case in the inviscid limit $\nu\to 0$ but
not at finite viscosity. So in turbulent flows, these probability distributions
have to be understood as the limit of their regularized analogue 
obtained by considering finite viscosity. For incompressible
and time-reversal invariant velocity fields
these probability distribution functions coincide.

The velocity probability distribution may be reconstructed
from the trajectory probability distribution functions.
For example, by expanding the trajectory equation (\ref{trajet}) to
lowest order, $x(t+\ep)=x(t)+\ep u(x(t),t)+\cdots$, one gets:
\debut
{\rm Proba}\(u(y,t)\in[v,v+dv]\)\, dv &=& \vev{\de(v-u(y,t))}\, dv 
\label{Uprob}\\
&=&\lim_{\ep\to 0^+}
P_{\rm backward}(y,t|y-\ep v,t-\ep)\, d\,(\ep\, v) \non
\fin
This formula applies only if the velocity is finite at each instant
and it assumes that trajectories are well-defined such that one may 
expand eq.(\ref{trajet}) to lowest order.

\vskip 0.3cm

\subsection{Kraichnan's passive scalar model.}
Definition, motivations and properties of the velocity fields,\\
Two-point function, energy cascade and inertial range.\\

In Kraichnan's model of passive advection,  \cite{kraichbis}, 
the statistics of the velocity field in eq.(\ref{eqT}), 
independent of the forcing, is supposed to be
gaussian with zero mean and with the two-point functions
\debut
\vev{u^a(x,t)u^b(y,t')}= D^{ab}(x-y)\de(t-t')
\quad {\rm with}\quad \nabla_{a}D^{ab}=0.
\label{vevu}
\fin
To analyze scaling properties we shall use the following expression for 
$D^{ab}$:
$D^{ab}(x)= D(0)\de^{ab} 
- d^{ab}(x)$ with
\debut
d^{ab}(x)= 
D\({ (d+\xi -1) \delta^{ab}- 
\xi \frac{x^a x^b}{|x|^2}}\)|x|^\xi
\label{defD}
\fin
where $\xi$ is a parameter, $0<\xi<2$. 
A more rigorous approach requires regularizing $D^{ab}(x)$
by introducing a infrared cut-off. For example:
$$
D^{ab}(x)=D_0\int dk\, \frac{e^{ik\cdot x}}{(k^2+m^2)^{(d+\xi)/2}}\,
\(\de^{ab}-\frac{k^ak^b}{k^2}\)
$$
Clearly, this distribution for $u$ is far from realistic.
It mimics however the growth of the correlations of 
velocity differences with separation distance, typical 
for turbulent flows since $\de u(x)\simeq r^{\xi/2}$.
The fact that the two-point functions 
(\ref{vevu}) are white noise in time 
is crucial for the solvability of the model
but it is of course very far from reality.
It in particular implies that, as in Brownian motion,
at each instant all the velocity moments
$\vev{u(x,t)^p}$ are infinite.

As for modeling turbulence,
the forcing term is also assumed to be gaussian with
mean zero and two-point function
\debut
\vev{f(x,t)f(y,t')}= C({\frac{_{x-y}}{^L}})\ \de(t-t')
\label{vevf}
\fin
The rotation-invariant function $C_L(x)=C(x/L)$, which could 
be chosen to be a gaussian, varies on scale $L$.

The parameter $\xi$ fixes the naive dimensions 
under rescalings $x\to \mu x$, $L\to\mu L$.  These are:
$$ 
[u] =\xi/2\quad ;\quad  [T] =(2-\xi)/2\quad;\quad [t]=(2-\xi)
$$  
The value $\xi=4/3$ corresponds to Kolmogorov scaling.
Indeed when comparing with Kolmogorov scaling one has to remember that
Kolmogorov typical turnover is finite and scales as $\tau_r\simeq r^{2/3}$
so that $\vev{(\de u)^2(r)}\simeq r^{2/3}\simeq r^{4/3} /\tau_r$.
This has to be compared with the two point function 
$\vev{(\de u)^2(r)}\simeq r^\xi\, \de(t)$ in Kraichnan's model.

Kraichnan's model provides a toy model for turbulence with a scale
domain, the inertial range, with constant energy flux. The scalar energy density
is $\CE=\half T^2$. In presence of non-zero molecular diffusivity,
the mean energy balance reads:
$$
\d_t\int \CE = \int \[ \vev{f\cdot T}- \kappa \vev{(\nabla T)^2} \]
$$
So, as in previous section, existence of a stationary regime requires
\debut
\bar \ep \equiv \kappa \vev{(\nabla T)^2} = \vev{f\cdot T} =\half C_L(0)
\label{balanT}
\fin
with $\bar \ep$ the mean dissipation rate. As in developped turbulence, one
has a dissipative anomaly since the mean dissipation rate $\kappa \vev{(\nabla T)^2}$
does not vanish in the limit $\kappa\to 0$.
Eq.(\ref{balanT}) simply means that the amount of energy transfered to the scalar
by the force equals the amount of energy dissipated.

Stationarity equation may be derived as in previous section,
see eqs.(\ref{2stat},\ref{station2}). Again let us do it in some details.
For example one may expand $T(x,t+\de t)$ 
to leading order and insert it into the correlations functions to derive the 
stationarity equations. Using eq.(\ref{sol1}) one has to expand 
the resolvant $R_\kappa$ which is formally define by the path ordered
exponential:
\debut
R_\kappa(x,t|y,t_0) = \[{P\cdot\exp[-\int_{t_0}^tds\, 
(\kappa \nabla^2 -\nabla\cdot u)(s)]}\]_{(x,y)}	\label{expR}
\fin
To leading order in $\de t$ this reads:
$$
R_\kappa(\cdot,t+\de t|\cdot,t)= 1 + \int\limits_t^{t+\de t}ds\, \nabla\cdot u(s)
- \kappa \nabla^2\, \de t 
+ \int\limits_t^{t+\de t}ds_1\int\limits_t^{s_1}ds_2\, 
(\nabla\cdot u)(s_1)\, (\nabla\cdot u)(s_2) + \cdots
$$
Recall that since both $f$ and $u$ are delta-correlated in time
the first term in the r.h.s. is of order $(\de t)^{1/2}$ but
the last term is of order $(\de t)$.
As a consequence one has for the two point functions:
\debut
&&\vev{T(x,t+\de t)T(y,t+\de t)} - \vev{T(x,t)T(y,t)} =
\kappa\({\nabla^2_x+\nabla_y^2}\)\vev{T(x,t)T(y,t)} \, \de t \non\\
&+&\int\limits_t^{t+\de t}ds_1\int\limits_t^{s_1}ds_2\,
\vev{(\nabla\cdot u)(x,s_1)(\nabla\cdot u)(x,s_2)T(x,t)T(y,t)} 
+ (x \leftrightarrow y)  \non\\
&+&\int\limits_t^{t+\de t}ds_1ds_2\,
\vev{(\nabla\cdot u)(x,s_1)(\nabla\cdot u)(y,s_2)T(x,t)T(y,t)}+
\int\limits_t^{t+\de t}ds_1ds_2\, \vev{f(x,s_1)f(y,s_2)} \non
\fin
All correlation functions may be factorized since by causality
$T(x,t)$ is independent of $u(y,s)$ or $f(y,s)$ for $s>t$;
eg. $\vev{u(x,s_1)u(y,s_2)T(x,t)T(y,t)}=\vev{u(x,s_1)u(y,s_2)}
\vev{T(x,t)T(y,t)}$ for $s_1,s_2>t$.

Assuming translation invariance, stationarity of $F_2(x)=\vev{T(x)T(0)}$
then gives:
\debut
\( -\kappa \nabla^2 - d^{ab}(x)\nabla_a\nabla_b\) F_2(x) =  C_L(x)
\label{hopf2}
\fin
The fact that it is a close equation reflects the linearity of the
transport equation and the particular choice of the velocity statistics.
This property makes Kraichnan's model theoretically attractive. 
In the rotational invariant sector the differential operator
$\CM_2^0\equiv -d^{ab}(x)\nabla^a\nabla^b$ is
$$
\CM_2^0 =-D(d-1)\, \inv{r^{d-1}}\frac{d}{dr}\, r^{d+\xi-1}\, \frac{d}{dr}
$$
With appropriate boundary conditions, the solution to eq.(\ref{hopf2}) is:
\debut
F_2(r)=\int_r^\infty d\rho\,  
\frac{\int_0^\rho du\,  u^{d-1}C_L(u) }{D(d-1)\rho^{d+\xi-1}+\kappa \rho^{d-1}}
\label{f2kraich}
\fin
It possesses an inviscid limit $\kappa\to0$ 
whose short distance expansion is:
\debut
F_2(r)_{\kappa=0}= {\rm const.} L^{2-\xi} -
\frac{2\bar \ep}{Dd(d-1)(2-\xi)}\, r^{2-\xi} + \cdots
\label{f0kraich}
\fin
The above ${\rm const.}$, which may be read of in eq.(\ref{f2kraich}), depends
on the details of the shape of the forcing correlation $C_L(x)$.
It is thus non universal. It however cancels in the two-point
structure functions
\debut
S_2(x) = \vev{(T(x)-T(0))^2} =
\frac{4\bar \ep}{Dd(d-1)(2-\xi)}\, r^{2-\xi} + \cdots
\label{s2kraich}
\fin
It is universal since it only depends on the distance and on the mean 
dissipation rate.
Note that we recover the naive scaling dimension of the scalar $[T]=(2-\xi)/2$.

To pursue further the analogy with the turbulent cascade, we should 
exhibit a constant energy transfer in some scale domain.
As in previous section, we introduce the energy dissipated $\CE_{\leq K}$
into modes of momenta $k$ less than $K$, as well as the energy $\CF_{\leq K}$
injected into these modes:
\debut
\CE_{\leq K} &=& \int_{|k|\leq K}\frac{dk}{(2\pi)^d}\ 
\int d^3x\, \kappa \vev{\nabla T(x) \nabla T(0)}e^{-ik\cdot x} \non\\
\CF_{\leq K}&=& \int_{|k|\leq K}\frac{dk}{(2\pi)^d}\ 
\int d^3x\, \vev{T(x)f(0)}e^{-ik\cdot x}\non
\fin
Fourier transforming eq.(\ref{hopf2}) then gives the energy
balance
$$
\CP_{\leq K}= \CF_{\leq K}-\CE_{\leq K}
$$
with $\CP_{\leq K}$ the energy flux through modes of momenta $k$, $|k|=K$.
Using the expression (\ref{f2kraich}) for $F_2$, one then verifies
that these quantities have the following $K$ dependence in the 
limit of zero molecular diffusivity, similar to the real turbulent case:

\vskip 1.0 truecm

$$\epsfbox{ 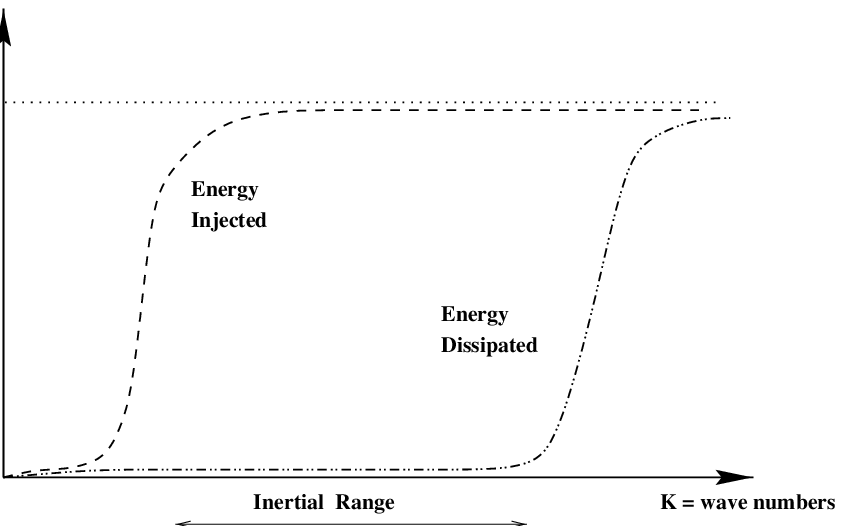}$$

\centerline{Figure 3: Scalar energy balance.}

\vskip 1.0 truecm

Hence the energy transfer $\CP_{\leq K}$ is
approximately constant in the domain $\eta\ll x \ll L$
with $\eta\simeq (\kappa/D)^{1/\xi}$ the dissipative scale
below which dissipation dominates over advection.
This is similar to fully developed turbulence.

\subsection{Anomalous scalings and universality.}
Two-point function with friction.\\
$N$-point functions and zero modes.\\
Anomalous scaling and universality.\\

If the scalar correlation functions were governed by a mean field theory,
the structure functions would scale as 
$\vev{(T(x)-T(0))^N}\simeq (\ep \bar r)^{N(2-\xi)/2}$.
Instead, we shall see that in the inertial range 
the even structure functions have anomalous scaling with \cite{allscalar}:
\debut
\vev{\({T(x,t)-T(0,t)}\)^N}\ \cong\ A_N
\({\frac{L}{|x|}}\)^{\rho_N}~|x|^{(2-\xi)N/2} +\cdots
\label{behav}
\fin
The anomalous exponents $\rho_N$ have been computed in 
a $\xi$-expansion or $1/d$-expansion \cite{allscalar,expo}:
\debut
\rho_N&=&\xi\frac{N(N-2)}{2(d+2)} + \CO(\xi^2) \label{anormal}\\
&=&\xi\frac{N(N-2)}{2d} + \CO(1/d^2) \non
\fin 
The exponents are universal depending only on 
$\xi$ but the amplitudes $A_N$ are not: they
depend on the shape of the covariance $C_L$.
These anomalous exponents are the signal of intermittency phenomenon,
with large fluctuations at short distances.

Let us now return to the problem of computing
the $N$-point correlation functions in Kraichnan's model. First we have to find
the constraints imposed by the stationarity condition
$$\d_t\vev{T(x_1)\cdots T(x_N)} = 0.$$
Using either ones of the methods explained in previous section 
leads to the following stationarity equations:
\debut
\CM_n^\kappa\, \vev{T(x_1)\cdots T(x_N)}
=\half\sum_{j,k}C_L(x_{jk})\, 
\vev{T(x_1)\cdots \hat {T(x_j)}\cdots \hat {T(x_k)}\cdots T(x_N)}
\label{kraichstat}
\fin
where overhatted quantities have to be omitted and,
assuming translation invariance,
\debut
\CM_N^\kappa
=-\kappa\sum_j\nabla_{x_j}^2 
+\half \sum_{j,k}d^{ab}(x_{jk})\nabla_{x_j}^a\nabla_{x_k}^b
\label{defMK}
\fin
Equations (\ref{kraichstat}) form a triangular set of equations which
recursively determine the $N$-point functions. In the inertial
range, scalar correlation functions satisfy eq.(\ref{kraichstat})
but with $\kappa\to 0$, ie. with 
\debut
\CM_N^0
=\half \sum_{j,k}d^{ab}(x_{jk})\nabla_{x_j}^a\nabla_{x_k}^b
\label{defMK0}
\fin 
This is a homogeneous singular second order differential operator
which possesses homogeneous zero modes that we shall denote $\varphi_n(x)$:
\debut
\CM_N^0\, \varphi_n(x) =0\quad,\quad
\varphi_n(\la x)= \la^{\xi_n}\, \varphi_n(x)
\label{zeromodes}
\fin
No explicit expressions for the zero modes are known, only $\xi$
or $1/d$ expansion have been found, see eq.(\ref{anormal}).

As in the case of the two-point function with friction described below, 
one can then show that these zero modes provide the dominating 
contributions to the inertial range correlation functions:
\debut
\vev{T(x_1)\cdots T(x_N)}= A_N\, \varphi_n(x)\, L^{\rho_N} + \cdots
\label{Nvev}
\fin
with $\rho_N =(2-\xi)N/2-\xi_N$. The dots refer 
to subleading terms as $L\to \infty$.
The anomalous scaling behaviors announced in eq.(\ref{behav}) 
then arise from these zero mode contributions. 

It is important to realize that universality of the scaling laws
(\ref{behav}) is ensured by the fact that $N$-point functions 
are dominated by zero modes.
The scaling dimensions only depends on the velocity field but not
on the force statistics, even if the latter would not be gaussian.

The large $N$ limit of the anomalous dimensions $\xi_N$ have been estimated 
numerically \cite{MazVer} or using instanton computations \cite{LebBal}.
One has:

$$ \epsfbox{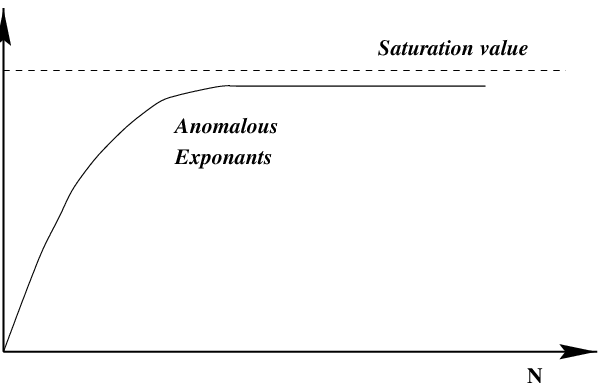}$$

\centerline{Figure 4 : Scalar anomalous dimensions.}

\vskip 1.0 truecm

So $\xi_N$ saturates as $N\to \infty$: $\xi_N\to \xi_\infty$.
Instanton calculations \cite{LebBal} give 
$\xi_\infty \simeq \frac{d(2-\xi)^2}{8\xi}$ for $d(2-\xi)\gg 1$.
In a multifractal description, see Section 2.2, this means 
that the minimum of the spectrum of fractal dimensions
for the variation of the scalar vanishes: $h_{\rm min}=0$.
Thus, there are subsets $\CV_0$ of the ambient space, with
${\rm dim}\, \CV_0=D_0$, on which the scalar varies discontinuously.
For typical velocity realization, the ambient space is decomposed
into domains, on which the scalar varies continuously, whose
boundaries $\CV_0$ are shock locations.

To illustrate the origin of these zero modes and anomalous scalings, let us
consider a simpler toy problem. It describes a passive scalar advected
by a smooth random velocity field but in presence of friction \cite{chertkov}. 
The friction is represented by an extra linear term, so that the
scalar equation of motion is:
\debut
\d_t T + \nabla\cdot(u\,T) +  T/\tau-  \kappa \De T = f.
\label{Tfric}
\fin
with $\tau$ the relaxation time induced by the friction.
The velocity is supposed to be gaussian with two-point
function (\ref{vevu},\ref{defD}) with $\xi=2$ such
the velocity correlations are smooth functions of the positions.
At $\kappa=0$ the stationarity equation for the two point function 
$F_2(x)=\vev{T(x)T(0)}$ reads:
\debut
-D(d-1)\inv{r^{d-1}}\(\frac{d}{dr}r^{d+1}\frac{d}{dr}\)F_2(r)
+\frac{2}{\tau} F_2(r) = C_L(r)
\non
\fin
This is solved by factorizing $F_2(r)=H(r)G(r)$ with $G$ 
solution of this differential equation without r.h.s..
This zero mode solution are homogeneous, $G(r)\propto r^{\al_\pm}$
with
\debut
\al_\pm = -\frac{d}{2} \pm \sqrt{\frac{d^2}{4} + \frac{2}{D(d-1)\tau}}
\label{alphapm}
\fin
With appropriate boundary conditions at infinity and at the origin, 
the result for $F_2$ is then:
$$
F_2(r) = r^{\alpha_+}\, 
\int_r^\infty\frac{d\rho}{D(d-1)\rho^{d+1+2\al_+}}
\int_0^\rho du u^{d+\al_+-1}C_L(u)
$$
At short distance, $r\to0$, it behaves as:
\debut
F_2(r) \simeq {\rm const.}\, r^{\al_+} +\cdots
\label{2pointfric}
\fin
Naive scaling analysis would have given scaling dimension
zero for $T$, since $\xi=2$. Instead the two-point function 
is dominated by a zero mode solution of stationarity equation
and as a consequence it acquires an anomalous scaling behavior.
Furthermore, the scaling exponent $\al_+$ is only function
of the velocity statistics and it is thus universal, whereas the
amplitude depends on the details of the force correlation and it 
is thus not universal.

More informations on the scalar turbulence problem 
may be found in \cite{plusscalar} and references therein.
I apologize for all possible omissions of this list.

\vfill \eject

\section{Lagrangian trajectories.}

\subsection{Richardson's law.}
Explosive separation of trajectories ($\not=$ chaos)\\
Breakdown of Lagrangian flows.\\

Richardson's law describes phenomenologically the separation 
of passive particles transported by turbulent flows \cite{richard}. 
Recall the definition (\ref{trajet}) of Lagrangian trajectories:
$$
\dot x(t)\ = \ u(x(t),t)
$$
with $u(x,t)$ the velocity field. 
Consider now two particles at positions $x_1$ and $x_2$
and let $\rho(t)=x_1(t)-x_2(t)$ be their relative position.
Assuming Kolmogorov's scaling, 
it satisfies in  the inviscid limit,
\debut
\dot \rho(t)  \simeq \bar \ep^{1/3}\, \rho^{1/3}    \label{rhopoint}
\fin
Thus, $\rho^{2/3}-\rho^{2/3}_0\simeq \bar\ep\, (t-t_0)$.
If the particles are at identical positions at $t_0=0$ ,
then
\debut
\rho^2(t)\ \simeq\ \bar \ep\, t^3 \label{richard}
\fin
This is Richardson's law: trajectories separate in time polynomialy. 
It describes the {\it explosive separation} of Lagrangian trajectories
in the limit of infinite Reynold's number,
ie. infinitely close trajectories separate in a finite time.

This behavior is different from properties of chaotic systems 
in which the separation of initially close trajectories is exponential. 
Indeed, suppose the trajectory equation would have taken the
form $\dot x= v(x,t)$ with $v(x,t)$ smooth enough such that its variation
on a scale $\rho$ is $\de v \simeq a\, \rho$. Then the analogue of eq.(\ref{rhopoint})
would have been $\dot\rho\simeq a\, \rho$ whose solution is exponential:
$\rho(t)\simeq \rho_0 e^{at}$. In particular, if trajectories start initially
at identical position, $\rho_0=0$, then they coincide at later time,
ie. $\rho(t)=0$. It is also clear from this tiny evaluation that
the explosive behavior (\ref{richard}) is linked to the non-smoothness
of the velocity field.

Eq.(\ref{richard}) means that two particles starting at 
positions which are indistinguishable in the inviscid limit, ie.
two initial positions distant less than the viscous scale $\eta$,
don't necessarily follow the same trajectory.
This statement has to be understood in a probabilistic sense since
eq.(\ref{richard}) is a mean field estimate for the averaged separation.
Richardson's law indicates that 
Lagrangian trajectories are ill-defined in the inviscid
limit since there may be different solutions to eq.(\ref{trajet})
with identical initial data. 
This explosive behavior may be rooted to the non-smoothness 
of the velocity field in the inviscid limit, since if it were smooth 
enough then Lagrangian trajectories (\ref{trajet}), which are
first order differential equations, would
be uniquely defined by their initial conditions.

\subsection{Lagrangian flows in Kraichnan's model.}
Path integral representation of the pdf's.\\
Heat kernel representation.\\

Recall equation (\ref{Tlagrange}) which links properties of the
scalar to those of the Lagrangian trajectories.
To take into account the diffusion of the passive scalar,  
we add a white-noise to the Lagrange equation and look for 
the statistics of trajectories defined by:
\debut
\dot x(t) = u(x(t),t) + \eta(t) 
\label{labruit}
\fin
with $u(x,t)$ a gaussian velocity field with statistics (\ref{vevu})
and $\eta$ a gaussian white-noise with two-point function,
$$
\vev{\eta(t)\eta(s)}= 2\kappa \, \delta(t-s)
$$
This modification will also take into account
possible ill-definedness of the probability distribution functions
of trajectories due to the non-smooth nature of the velocity field
in Kraichnan's model.

Since $u(x,t)$ is time-reversal invariant, the backward and forward 
probability distribution functions coincide. 
They admit a simple representation as a heat kernel for the
differential operator $\CM_n^\kappa$:
\debut
P^{(N)}_\kappa(x,t|x_0,t_0) = \Bigl[\exp\(-|t-t_0|\CM^\kappa_N\)\Bigr]_{(x|x_0)}
\label{Pheat}
\fin
with $\CM_n^\kappa$ defined in eq.(\ref{defMK}).
This representation of $P^{(N)}(x,t|y,s)$ together with eq.(\ref{Tlagrange})
provides another way to derive the stationarity equation (\ref{kraichstat}).

A possible derivation of eq.(\ref{Pheat}) uses path integral manipulation.
By introducing a path integral representation of
the delta-function coding the trajectories, one has:
\debut
P^{(N)}_\kappa(x,t|x_0,t_0) &=&
\vev{\prod_{j=1}^N \de(x_j- x(t_j|x_{0\,j},t_{0\, j}))} \label{Pkrach}\\
&=&\vev{\prod_{j=1}^N \int\limits^{x(t)=x}_{x(t_0)=x_0} Dp\, Dx\,
e^{-i\int_{t_0}^tds\, p_j(s)(\dot x_j(s) -u(x_j(s),s) - \eta(s))} }\non
\fin
The average is over $u$ and $\eta$. Since they are both
gaussian we get:
\debut
&& ~~~~~~~~~~~~~~~~~~~~ P^{(N)}_\kappa(x,t|x_0,t_0)  = \label{Pkrach2}\\
&&\int\limits^{x(t)=x}_{x(t_0)=x_0} Dp\, Dx\
\exp{\[ -\int_{t_0}^tds\(\sum_j(\kappa p^2_j(s)+i p_j(s)\cdot x_j(s)) 
+\half\sum_{j,k} D^{ab}(x_j-x_k)p^a_j(s)p^b_k(s)\)\] }
\non
\fin
This is the path integral representation of the heat kernel of $\CM_N^\kappa$.

In absence of white-noise in Lagrange equation, 
i.e. in the limit of $\kappa\to 0$, we get:
\debut
P^{(N)}_{\kappa=0}(x,t|x_0,t_0) = \Bigl[\exp\(-|t-t_0|\CM^0_N\)\Bigr]_{(x|x_0)}
\label{Pheat2}
\fin
with $\CM^0_N$ the $\kappa=0$ limit of $\CM_N^\kappa$ defined in eq.(\ref{defMK0}).
For $\xi=0$, $\CM^0_N$ reduces to the Laplacian operator and $P^{(N)}_{\kappa=0}$
to the usual heat kernel of the Brownian motion.

The kernel $P^{(N)}_{\kappa=0}$ satisfy the semi-group law:
$$
\int dy P^{(N)}_{\kappa=0}(x,0|y,s)P^{(N)}_{\kappa=0}(y,s|z,t)
=P^{(N)}_{\kappa=0}(x,0|z,t)
$$
They thus define a Markov process.
Translation invariance ensures that $P^{(N)}$ only depends on the time difference.
Time-reversal invariance of the velocity statistics implies that
$P^{(N)}(x,t|y,s)=P^{(N)}(y,t|x,s)$.
Due to the homogeneity property of $\CM_N^0$ we have
$$
\la^{Nd}\,P^{(N)}_{\kappa=0}(\la x,\la^{(2-\xi)}t|\la x_0,\la^{(2-\xi)}t_0)
=P^{(N)}_{\kappa=0}(x,t|x_0,t_0)
$$

\subsection{Slow modes.}
Hyperdiffusivity.\\
zero modes as conserved quantities and slow modes.\\

The zero modes, which dominate the scalar correlations in the
inertial range, have a simple interpretation 
in terms of Lagrangian trajectories:
{\it They are quantities conserved in average.}

One may test the distribution of the trajectories by computing 
averages of some functions $f(x)$ of the particle positions:
\debut
\vev{f}_{(t|x_0,t_0)}\equiv \int dx\, P^{(N)}_{\kappa=0}(x,t|x_0,t_0)\, f(x)
\label{ave}
\fin
This is a function of the initial positions and of the time $t$ at which
we evaluate it. To avoid irrelevant technical problems we shall only consider
translation invariant test functions.

Generically, these averages grow with time indicating
an increase of the distance between the particles. More precisely,
if the test function is homogeneous of degree $\sig$,
$f(\la x)= \la^\sig f(x)$, then
\debut
\vev{f}_{(t|x_0,t_0)} \simeq \, t^{\sig/(2-\xi)}
\label{hyperdiff}
\fin
When $\xi=0$ one recover the usual diffusive behavior of the Brownian motion.
For $\xi<2$ the increase is faster and one usual refers to it as super-diffusive.
Note that the exponents diverge as $\xi\to 2$.
Eq.(\ref{hyperdiff}) follows from homogeneity property 
of the heat kernel $P^{(N)}_{\kappa=0}$.

The zero modes are characterized by completely atypical behavior.
If $\varphi_n(x)$ is a zero mode, $\CM_N^0\varphi_n=0$, then:
\debut
\vev{\varphi_n}_{(t|x_0,t_0)} = \varphi_n(x_0),\quad
{\rm independent ~ of ~ time}
\label{conserve}
\fin
So the zero modes reflect {\it coherent structures} 
which are preserved by the flow.
The derivation of eq.(\ref{conserve}) is
simple. By definition (\ref{ave}), one has:
\debut
\d_t \vev{\varphi_n}(t|x_0,t_0) &=& \int dx\, \d_t
P^{(N)}_{\kappa=0}(x,t|x_0,t_0)\, \varphi_n(x) \non\\
&=& - \vev{ (\CM_N^0\varphi_n)}(t|x_0,t_0) =0 \non
\fin
where we used that $(\d_t+\CM_N^0)P^{(N)}_{\kappa=0}(x,t|y,s)=0$.

Besides the conserved zero modes $\varphi_n(x)$, one actually
has a series of modes, that we denote as
$\varphi_{n;k}(x)$, $k\geq 0$, whose correlation functions increase
slower with time than the hyperdiffusive law (\ref{hyperdiff}).
They are related by:
\debut
\CM_N^0\, \varphi_{n;k+1}(x) = \varphi_{n;k}(x)
\label{descente}
\fin
They form a tower of descendants whose top is the zero modes
$\varphi_n=\varphi_{n;0}$.
In particular, if the zero mode has dimension $\xi_n$ then
its descendants $\varphi_{n;k}$ has dimension $\xi_{n;k}=
\xi_n + (2-\xi)k$.
These modes code for the behavior of the Lagrangian trajectories
as points become closer. More precisely the probability
distribution functions of the trajectories admit a short
distance expansion of the following form:
\debut
P^{(N)}_{\kappa=0}(x,t|\la y,0) \simeq
\sum_{n;k}\la^{\xi_{n;k}}\,\psi_{n;k}(x,t)\, \varphi_{n;k}(y)
\label{expans}
\fin
as $\la \to 0$.
The descent equation (\ref{descente}) follows from consistency
conditions for this expansion. Indeed inserting the defining relation 
$(\d_t+\CM_N^0)P^{(N)}_{\kappa=0}(x,t|y,0)=0$ into this expansion 
implies $ (\d_t+\CM_N^0)\psi_{n;k}(x,t)=0$. Similarly, plugging
the expansion (\ref{expans}) into $(\d_tP^{(N)}_{\kappa=0}
+P^{(N)}_{\kappa=0}\CM_N^0) (x,t|y,0)=0$ yields:
$$\sum_{n;k}\la^{\xi_{n;k}}\,\d_t\psi_{n;k}(x,t)\, \varphi_{n;k}(y)
= -\sum_{n;p}\la^{\xi_{n;p}+(\xi-2)}\,\psi_{n;p}(x,t)\, (\CM_N^0\varphi_{n;p})(y)
$$
where we used the fact that $\CM_N^0$ is homogeneous with scaling
dimension $(\xi-2)$. Comparing the left and right hand sides
shows that the scaling dimensions $\xi_{n;p}$ should gather into
families with $\xi_{n;p}=\xi_{n;0}+p(2-\xi)$ with the modes
$\varphi_{n;p}$ related by the descent equation (\ref{descente}) and
$\d_t\psi_{n;p}=-\psi_{n;p+1}$. The modes $\varphi_{n;p}$ are
called {\it slow modes} because their expectation values,
as defined in eq.(\ref{ave}), increase with time slower than
what the generic hyperdiffusive (\ref{hyperdiff}) behavior would have predicted.
Expansion (\ref{expans}) may be interpreted as a 
kind operator product expansion.

\subsection{Breakdown of Lagrangian flows.}
Limit of coincident initial points. \\
Ill-definiteness of trajectories.\\

Let us now see what happen to $P^{(N)}_{\kappa=0}$ as initial points are 
approaching each others. As the velocity field is time reversal
invariant, this is equivalent to approaching the final points.
If the Lagrangian trajectories were well-defined, ie. if the
Lagrangian trajectories were uniquely defined by their initial
or final positions, we would have:
\debut
P^{(N)}(x,t|y,s)\Big\vert_{y_1=y_2}
= \delta(x_1-x_2)\, P^{(N-1)}(x,t|y,s)
\label{welldef}
\fin
Instead for $\xi<2$ the limiting distribution, 
\debut 
P^{(N)}_{\kappa=0}(x,t|0,0) = \[\exp(-t\CM_N^0)\]_{(x|0)}
\label{Plimit}
\fin
is a regular function, ie. it is not concentrated on the delta function.
This may be checked in the limit $\xi\to 0$ in which $\CM_N^0$
becomes the Laplacian operator. These functions  give the probability for
trajectories starting at the same initial point to split. 
A more precise description of this splitting is provided 
by the expansion (\ref{expans}).

The distribution of two particles starting at identical initial
positions may be computed exactly in the rotation invariant
sector. After changing variable from $r$ to $z=r^{(2-\xi)/2}$, 
the differential operator $\CM^0_2$ becomes
$$
\CM_2^0 = D'\, z^{-h}\[{-(\frac{d}{dz})^2 + \frac{h(h-1)}{z^2}}\]z^h
$$
with $h=\frac{d}{2-\xi}-\frac{1}{2}$ and $D'=D(d-1)$. The distribution of 
two particles $P^{(2)}_{\kappa=0}(z,t|0)$ is then solution
of $(\d_t+\CM_2^0)P^{(2)}_{\kappa=0}=0$ with the boundary condition
$P^{(2)}_{\kappa=0}(z,t=0|0)\, z^{2h}dz = dz\, \de(z)$. The result is:
\debut
P^{(2)}_{\kappa=0}(r,t|0)\, r^{d-1}dr = {\rm const.}\,
(\frac{z^2}{t})^h\, \exp\({-z^2/4D't}\)\, \frac{dz}{t^{1/2}}
\label{2pointidem}
\fin

The violation of equation (\ref{welldef}) implies a breakdown
of the Lagrangian flow in Kraichnan's velocity fields.
This breakdown is linked to the irregularities of the velocity;
irregularities which imply the existence of many solutions to 
the Lagrange equation (\ref{trajet}). It is similar to the one
predicted by Richardson's law.
A meaning to the distribution of these solutions at fixed 
velocity realization has been given in \cite{matheux}.

This spreading of the trajectories implies an information loss,
as knowing that two particles are at identical positions at time $t_0$
does not guarantee that this information will be true at later time.
A more quantitative formulation of this information loss will
be welcome.


\subsection{Batchelor limit.}
Trajectory pdf's for smooth fields.\\
Trajectories are well-defined.\\
Lyapunov exponents and universality.\\

The Batchelor limit is the limit of smooth velocity field,
ie. $\xi=2$ in Kraichnan's model.
In this limit Lagrangian trajectories have different behaviors,
closer to those of trajectories in chaotic systems.

The probability distribution function for two trajectories 
satisfies $(\d_t+\CM_2^0)P^{(2)}=0$ which for $\xi=2$ and
in the rotation and translation invariant sector reduces to:
\debut
\({\d_t - \frac{D'}{r^{d-1}}\frac{d}{dr}r^{d+1}\frac{d}{dr} }\) 
P^{(2)}(r,t|r_0,0) =0 
\non
\fin
with the boundary condition $r^{d-1}P_2(r,0|r_0,0)\, dr=\de(r-r_0)\, dr$.
The solution is:
\debut
P^{(2)}(r,t|r_0,0) = \frac{r^{-d}}{\sqrt{4\pi D't}}
\exp\({-\inv{4D't}\(\log(r/r_0)-dD't\)^2}\) \label{pdeux}
\fin
with $D'=D(d-1)$. Its limit of initial coincident points is such that:
$$
\lim_{r_0\to 0}\, P^{(2)}(r,t|r_0,0)\,r^{d-1}dr = \de(r)\, dr
$$
Thus eq.(\ref{welldef}) is satisfied meaning that trajectories
are well-defined. This could have been expected since for
$\xi=2$ Kraichnan's velocity field is smooth
because the typical value of its variation between
points distant by $r$ is $\de u \simeq r$.

However the trajectories separate exponentially
as it can be seen from eq.(\ref{pdeux}) 
or from eq.(\ref{hyperdiff}) in the limit $\xi\to 2$.
This exponential separation prevent us to use the Batchelor limit 
to cure the ill-definedness of Lagrangian trajectories.
Indeed assume that the turbulent velocity fields have been 
regularized  such that they become smooth on scales less than say $\eta$
but  remain irregular above this scale. 
Lagrangian trajectory whose initial points are 
distant less than $\eta$ first enjoy an exponential
separation until they are distant by a length of order $\eta$,
and then follow different trajectories since they are well separated.

In this smooth limit, trajectory statistics is solvable by reducing
it to a group theoretical problem \cite{shrai,slow}.
Let us just illustrate this point by computing the Lyapunov
exponents of the Lagrangian flows \cite{lyapu}.
Since the velocity is smooth, the velocity variation may be parameterized as
$\de u^a(x,t)=\sig^{ab}(t) x^b$ with $\sigma(t)$ a traceless $d\times d$ 
matrix, $\sigma(t)\in sl(d)$. For the Kraichnan's model
$\sigma(t)$ is gaussian with two-point function:
$$
\vev{\sig^{ab}(t)\, \sig^{cd}(s)}
=2D\((d+1)\de^{ac}\de^{bd}-\de^{ab}\de^{cd}-\de^{ad}\de^{bc}\)\, \de(t-s)
\equiv C^{ab;cd}\, \de(t-s)
$$
The equation for the separation of Lagrangian trajectories is then:
\debut
\dot x(t) = \sigma(t)\cdot x(t) \label{trajetreg}
\fin
Its formal solution is
\debut
x^a(t) = G^{ab}(t)\, x^b_0
\quad ,\quad G(t) = P\exp\(\int_{t_0}^t ds \sigma(s)\) \label{Wint}
\fin
The matrix $G(t)$, which belongs to the group $SL(d)$, $\det G=1$,
codes the information for the Lagrangian flows.
The later may thus be seen as a process on the group $SL(d)$
with 
\debut
\dot G(t) = \sig(t)\, G(t) \label{trajetgroup}
\fin
It is a Markov process which is fully
determined by the probability transition $P_t(G,G_0)$
from the group element $G_0$ to $G$. It is formally defined
as $P_t(G,G_0)=\vev{\de(G^{-1}G(t,t_0))}$ with $\de(\cdot)$ the Dirac measure
for the Haar measure on $SL(d)$ and $G(t,t_0)$ solution of eq.(\ref{trajetgroup})
with initial condition $G_0$.
The associated Fokker-Planck equation, which may be derived using 
techniques explained above for the Navier-Stokes equation, reads:
\debut
\({\d_t - \half C^{ab;cd}\, L_{cd}\,L_{ab}}\) P_t(G,G_0)=0
\label{batchFP}
\fin
with the initial condition $P_t(G,G_0)\vert_{t=0}=\de(G^{-1}G_0)$.
Here $L_{ab}$ are the vector fields corresponding to 
the infinitesimal left-action of $gl(d)$ on $d\times d$ matrices,
$L_{ab}f(G)= \frac{d}{dt}f(e^{tE_{ab}}G)\vert_{t=0}$ with
$E_{ab}$ the elementary matrices.
The Fokker-Planck equation (\ref{batchFP}) may alternatively be 
written as sum of Casimir operators:
$$\({\d_t - D(d+1) J^2 -DdH^2}\) P_t(G,G_0)=0$$ 
with
$H^2=\sum_{ab} \hat L_{ab}\hat L_{ba}$ and
$J^2=\half\sum_{ab} (\hat L_{ab}-\hat L_{ba})^2$ where
$\hat L_{ab}=L_{ab}-\inv{d}\sum_aL_{aa}$
are generators of $sl(d)$.

Rotations of the initial (final) points correspond respectively
to the right (left) actions of $SO(d)$ on $SL(d)$.
So observables invariant by initial and final rotations 
are functions on the coset space $SO(d)\backslash SL(d)/SO(d)$.
Examples of functions on this space are provided by traces
${\rm tr}(GG^T)$ in any representation of $SL(d)$.
It is convenient to Iwasawa decompose $G$ as $G=O_1N$ with $O_1\in SO(d)$
and $N$ a lower triangular matrix. The later may be factorized
as $N=DO_2$ with $O_2\in SO(d)$ and $D$ diagonal,
$D={\rm diag}(z_1,z_2,\cdots,z_d)$ with $\prod_j z_j=1$.
This identifies points of the coset space $SO(d)\backslash SL(d)/SO(d)$
with diagonal matrices.
The eigenvalues $z_j$ describe how blobs of initial particles
are stretched. They typically increase exponentially.
So that at large time the logarithms of these eigenvalues
are asymptotically linear in time:
$$\log z_j \simeq \la_j t \quad {\rm as}\quad t\to \infty$$
The coefficient $\la_j$ are called the Lyapunov exponents.

On $SO(d)\backslash SL(d)/SO(d)$, the Fokker-Planck equation (\ref{batchFP})
reduces to:
\debut
\({\d_t - Dd\sum_j\d_{y_j}^2 + D(\sum_j\d_{y_j})^2
- Dd\sum_{i\not= j}\coth(y_i-y_j)\d_{y_j}}\) 
P_t(y;y_0)=0
\fin
with $z_j=\exp y_j$, $\sum_j y_j=0$. 
This is the Schroedinger equation
for a Calogero-Sutherland type hamiltonian.
It is also the Fokker-Planck equation for the Langevin dynamics:
\debut
\d_t y_j = Dd\sum_{i\not= j}\coth(y_i-y_j) + \xi_j 
\label{calonoice}
\fin
where $\xi_j$ are random gaussian variables with zero mean and
$\vev{\xi_j(t)\xi_i(s)}=2D(d\de_{ij}-1)\de(t-s)$.
The typical dynamics is simple to analyze. Suppose
that at initial time the eigenvalues
are ordered $\la_1>\la_2>\cdots>\la_d$. 
The interacting term in (\ref{calonoice}) preserves this
order and increases the splitting between the eigenvalues.
So at large time $\la_1\gg \la_2\gg \cdots \gg \la_d$
and we can then approximate $\coth(y_i-y_j)$ by $\pm 1$.
The effective dynamics of the eigenvalues then reads:
$$\d_t y_j = Dd\,(d-2j+1) + \xi_j$$
This shows that the Lyapunov exponents are 
$\la_j=Dd\,(d-2j+1)$. This effective dynamics 
also shows that at large time the statistics of
the logarithm of the eigenvalues is gaussian.

Trajectories in smooth but time-correlated velocity fields
have been analyzed in \cite{balko}. The Lagrange equations are
then eq.(\ref{trajetreg}), or equivalently eq.(\ref{trajetgroup}),
but with $\sig(t)$ time-correlated. Decomposing the group element
as above $G=ON$ with $O\in SO(d)$ and $N$ triangular, the time evolution reads 
$O^{-1}\dot O + \dot N N^{-1} = O^{-1} \sig(t) O$.
Let as above $z_j=\exp y_j$ be the eigenvalues of $N$, then
$\dot z_j z_j^{-1} = (O^{-1} \sig O)_{jj} \equiv \hat \sig_{jj}$.
This is simply integrated as:
$$
y_j(t) = y_{0 j} + \int_0^t ds\, \hat \sig_{jj}(s)
$$
If $\tau$ is the correlation time of $\hat \sig$, the above integral
is a sum of $t/\tau$ independent variables. We can apply the 
central limit theorem for $t$ large. The bulk of the distribution
of the $y_j's$ is thus gaussian with mean average increasing linearly
with time. The proportionality coefficient $\la_j$ are the Lyapunov exponents
with
$$
\la_j = \inv{\tau} \int_0^\tau ds\, \vev{\hat \sig_{jj}(s)}
$$
The gaussianity of the statistics illustrates the fact that the
statistics of Lagrangian trajectories in time-correlated but smooth
velocity fields are described by effective universal dynamics.
See ref.\cite{balko} for a discussion of consequences of this property
on scalar turbulence.

\subsection{Generalized Lagrangian flows and trajectory bundles.}
Non-uniqueness of trajectories and parameters as random variables.\\
Generalized Lagrangian flows.\\

From the two preceding sections we learn that Lagrangian trajectories
are well-defined for $\xi=2$, ie. when the velocity field is regular,
but ill-defined for $\xi<2$ in the sense that eq.(\ref{welldef})
is broken. This arises from the fact that there is many solutions
to the equation $\dot x = u(x,t)$ if $u(x,t)$ is not
smooth enough. To exemplify this fact consider $\dot x= x^\sig$ with 
$0<\sig<1$ which possesses two solutions with initial conditions
$x(0)=0$, namely $x(t)\propto t^{1/1-\sig}$ or $x(t)=0$.
So the probability distribution $P^{(N)}$ are actually not describing
discrete set of particles but diffuse clouds of trajectories.

More generally, let us parameterize the set of solutions
to Lagrange equation as $x(\om,t)$ so that
\debut
\dot x(\om,t)=u(x(\om,t),t)	\label{general}
\fin
with $u(y,t)$ some realization of the velocity field.
One may thought of $\om$ as labeling the initial condition plus 
the extra parameters needed to specify the solutions. To make it more
precise, but more formal, one assumes \cite{brenier,shnirel} that
$\om\in \Om$ with $\Om$ a space of events equipped with a probability law,
ie. equipped with a measure $\mu(d\om)$. The data of map
$x:\Om\times [t_i,t_f]\to R^d$ whose output is $x(\om,t)$ has been called
generalized Lagrangian flow.

Incompressibility may be forced by demanding 
that the measure $\mu(d\om)$ satisfies:
\debut
\int \mu(d\om)\, \varphi_t(x(\om,t))= \int dy\, \varphi_t(y)
\label{incom}
\fin
for sufficiently regular functions $\varphi_t(y)$ on $R^d$.
Indeed time differentiating eq.(\ref{incom}) leads to
the weak incompressibility condition 
$\int dy \,u(y,t)\cdot\nabla\varphi_t(y)=0$.

The case of well-defined trajectories corresponds to $\Om=R^d$
with $d$ the dimension of the ambient space and $\mu$ the
Lebesgue measure. The parameter $\om$ is then the initial position.
In the opposite case with a breakdown of the
Lagrangian flow we should talk not about single trajectory but
only about what may be called a trajectory bundle
with some probability distribution law, see figure.

\vskip 1.0 truecm

$$ \epsfbox{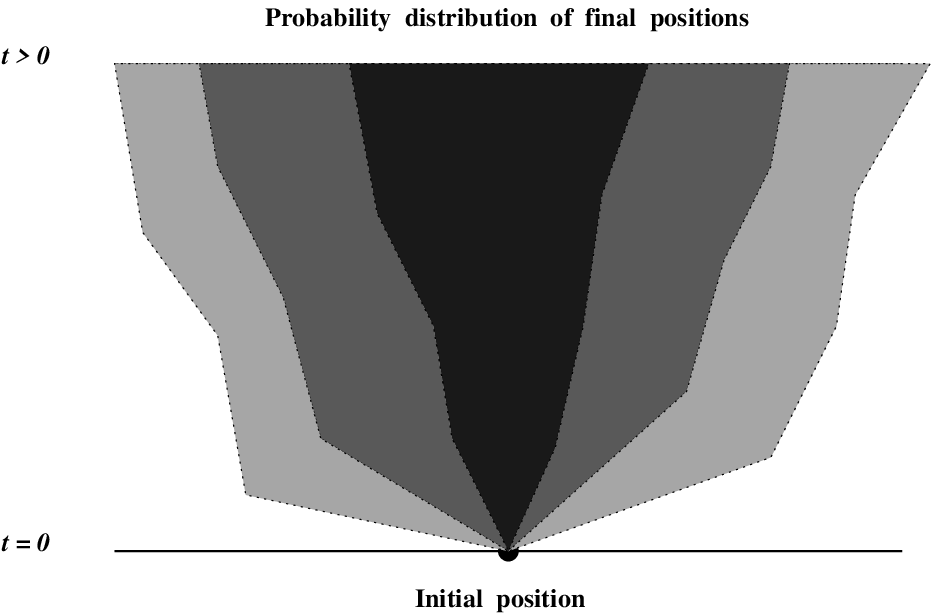} $$

\centerline{ Figure 5 : Trajectory bundles.}

\vskip 1.0 truecm 

For example the probability to find a trajectory in a domain $\CD \subset R^d$
a time $t$ knowing that it was inside a domain $\CD_0 \subset R^d$
a time $t_0$ is given by:
\debut
P_{t,t_0}(\CD,\CD_0|u)= \int_\CD dx\,\int_{\Om_{\CD_0,t_0}}\mu(d\om)\, 
\de(x-x(\om,t))
\label{generalP}
\fin
wih $ \Om_{\CD_0,t_0}=\{\om\in\Om\ {\rm s.t}\ x(\om,t_0)\in\CD_0\}$
the subset of events with trajectories initially inside $\CD_0$.
It is worth stressing that this is a probability distribution
at fixed velocity realization.

In the case of Kraichnan's model this probability
distribution for a single trajectory bundle is
$$
P_{t,t_0}(\CD,\CD_0|u)= \int_\CD dx \int_{\CD_0}dy\, R_{\kappa=0}(x,t|y,t_0)
$$
with $R_{\kappa=0}(x,t|y,t_0)=\[P\cdot \exp\int_{t_0}^t (\nabla\cdot u)(s)\]$
defined in eq.(\ref{defR},\ref{expR}). This is still a random function as
Kraichnan's velocity is itself random. The average of it gives the probability
distribution $P^{(N=1)}_{\kappa=0}(x,t|y,t_0)$.

It will be quite interesting to extend the previous analysis
(conserved and slow modes, ill-definedness and
probability distribution functions of trajectories, universality, etc...)
to non-smooth and time-correlated velocity fields.

\vfill \eject

\section{Burgers turbulence.}
Burgers equation and shocks.\\
Bifractality and velocity pdf's.\\

The forced Burgers equation in $1+1$ dimension is the following:
\debut
\d_t u + u\d_x u -\nu \d_x^2 u = f \label{burgers}
\fin
The force is supposed to be gaussian with covariance
$\vev{f(x,t)f(y,s)}=C(x-y)\de(t-s)$.
There is no pressure and thus no incompressibility condition is
imposed on the velocity field $u(x,t)$. Although its properties are
very different from those of the Navier-Stokes equation both
equations possess the same non-linearity. The Burgers equation 
hence provides an interesting toy model for turbulence
in which progress have recently been done, 
cf eg \cite{buug,EvE} and references therein.
The main point about Burgers turbulence is that one knows the
geometrical structure responsible for intermittency:
these are shocks.

Solutions of eq.(\ref{burgers}) developp shocks in the
inviscid limit. The occurence of these shocks are esay to 
understand by considering the Burgers equation in absence of force. 
In the inviscid limit it reduces to the Euler
equation $\d_t u + u\d_x u=0$ whose solutions are such that
$u(x,t)=u_0(x-tu(x,t))$. Assume for example that the initial condition
$u_0(x)$ is such $x=-\al u_0^3(x)-\tau u_0(x)$ with $\al$ and $\tau$
positive. Then at latter time
$u(x,t)$ satisfies $x=-\al u^3(x,t)-(\tau-t) u(x,t)$.
Hence  $u(x,t)$ would be multivalued for $t>\tau$ which means
that a shock has developped at time $\tau$.

The Burgers equation may be rewritten in different ways.
First setting $u=\d_xh$ and $f=-\d_xV$ it becomes the
so-called KPZ equation : $\d_t h+\half(\d h)^2 - \nu \d^2h=-V$.
Furtheremore, setting $h=-\inv{2\nu}\log\psi$, it becomes equivalent
to the one dimensional Schrodinger in a random potential:
$\d_t\psi= \[-\d^2+2\nu V\]\psi$.

For typical realization the velocity profile will be 
piecewise smooth, meaning that the profile is made of
a succession of intervals, on which the velocity
is smooth, these intervals being separated by shocks.


$$\epsfbox{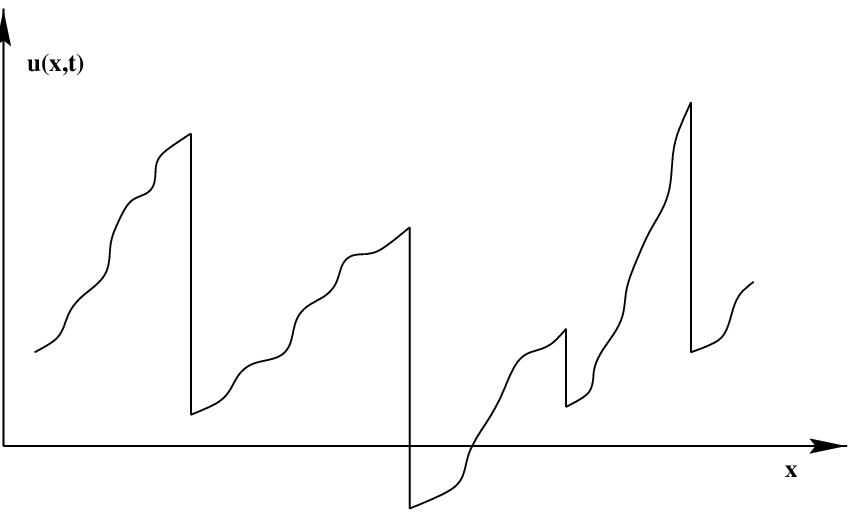}$$

\centerline{Figure 6 : Typical profile in Burger's trubulence.}

\vskip 1.0 truecm

More explicitly as a function of the position the velocity
is discontinuous so that its derivative may be decomposed as:
\debut
\d_xu(x,t) = \xi(x,t) + \sum_s a_s(t) \de(x -x_s(t) )
\label{deru}
\fin
The sum is over the shocks, which are known to form
a discrete set. The points $x_s$ are the
shock locations and $a_s$ their amplitudes.
The function $\xi(x,t)$ is regular except at the points $x_s$ where it
may have discontinuities. 
So $\d_x\xi(x,t)$ may also have delta function singularities.

The positions and the amplitudes of the shocks are time
dependent. The velocity of the shocks is the mean value
of the velocity on the two side of the shock so that
\debut
\dot x_s &=& v_s \quad {\rm with}\quad  u(x_s^\pm,t)=v_s \pm a_s/2
\label{uchoc}
\fin
By definition the amplitude of the shock is $a_s=u(x_s^+)-u(x_s^-)$.
The shock dynamics follows directly from Burgers equation.
Time derivating eq.(\ref{uchoc}) yields
$\d_t(v_s\pm \frac{a_s}{2})=\dot x_s(\d_x u)(x^\pm_s) +(\d_t u)(x_s^\pm)$.
But outside shocks there are no anomaly in the inviscid limit and
the inviscid Burgers equation $\d_tu+u\d_xu=f$ holds.
Applying it on the two sides of the shocks, ie. for $x\to x_s^\pm$, gives 
$(\d_t u)(x_s^\pm) + (v_s\pm \frac{a_s}{2})(\d_x u)(x^\pm_s) = f(x_s)$.
By definition $(\d_x u)(x^\pm_s)=\xi(x^\pm_s)$ so that:
\debut
\dot v_s &=&-\frac{a_s}{4}(\xi(x_s^+)-\xi(x_s^-)) + f(x_s) \non\\
\dot a_s &=&-\frac{a_s}{2}(\xi(x_s^+)+\xi(x_s^-)) \non
\fin
Time evolution of higher derivatives of the velocity at a shock may be
derived in a similar way. Namely, let $\xi^\pm_n = \inv{n!}(\d^n u)(x_s^\pm)$
and $f_n=\inv{n!}(\d^n f)(x_s)$, then applying Burgers equation on the
two sides of the shock gives:
$$
\dot \xi^\pm_n - (n+1)\, v_s\, \xi^\pm_{n+1} 
+\sum_{p+q=n+1}p \xi^\pm_p\xi^\pm_q=f_n
$$

At finite viscocity there is no shock, ie. no discontinuity
of the velocity. The shocks are becoming domains of width
of order $1/\nu$ over which the velocity decreases rapidly.
Inside the  shock domain located at $x_s$ the velocity will be
of the form
\debut
u_s(x,t) = w((x-x_s)/\nu,t;\nu)
\quad {\rm with}\quad w(z,t;\nu)= w_0(z,t)+ \nu w_1(z,t)+\cdots
\label{danschoc}
\fin
The condition to connect this solution to the rest of the velocity
profile on the two sides of the shock is that $w(\pm\infty)=u(x_s^\pm)$.
Solving the Burgers equation with this ansatz gives to
lowest order 
\debut
w_0(z)=v_s-\frac{a_s}{2}\tanh(a_sz/2) \label{w0shock}
\fin

This knowledge may for example be used to illustrate the presence
of dissipative anomalies. Consider the dissipative field
$\ep(x)=\frac{\nu}{2}(\d_xu)^2$. For a given realization,
$u(x)$ is smooth outside shocks and $\nu(\d_xu)^2$ vanishes
there. Thus only shocks contribute to the dissipation.
To find the contribution of a given shock, localized at $x_s$,
one uses the representation (\ref{w0shock}) of the velocity inside
the shock to compute the dissipation. 
One has $\frac{\nu}{2}(\d_xu_s)^2 \simeq \frac{|a_s|^3}{12}
\de(x-x_s)$ as $\nu\to 0$, so that $\ep(x)$ is the sum of the
shock contributions:
\debut
\ep(x)=\sum_{s:{\rm shock}}\frac{|a_s|^3}{12} \de(x-x_s) \label{diss}
\fin
More generally, one may also consider family of operators
defined as products of the dissipation
by functional of the velocity, eg. $\ep(x)\, \de(v-u(x))$,
or consider higher derivatives of the velocity.
They are all localized on shocks and they
code for the structure of the velocity
profile inside and around the shocks.

Shocks dominate many properties of Burgers turbulence.
The first consequence is that the velocity structure functions
are bifractal so that Burgers turbulence manifests a pronounced 
intermittent behavior. One has for $x$ small and positive:
\debut
\vev{(\de u)^p(x)} = \cases{ \vev{\xi^p}\, x^p, &for $0<p<1$;\cr
	\rho_s \vev{a_s^p}\, x, & for $1<p$.\cr}
\label{fractalburg}
\fin
with $\rho_s$ the shock density.
In other words, for $0<p<1$ the moments are dominated by the smooth
parts of the velocity, whereas for $1<p$ they are dominated by
the shocks. This may be understood as follows.
If there is no shock in the interval $[y,y+x]$, for small $x$
the variation  of the velocity between 
the two points $y$ and $y+x$
is $\de u(x) \simeq\xi\, x$ with $\xi$ the derivative of $u$. 
The probability for such an event is $(1-\rho_s x)$ for small $x$.
If there is a shock in the interval $[y,y+x]$, the velocity variation is 
$\de u(x)\simeq a_s$ with $a_s$ the amplitude of the shocks. The probability
for this event is $\rho_s x$. Thus the moments may be evaluated as:
$$
\vev{(\de u)^p(x)} \simeq (1-\rho_s x)\, \vev{(\xi\, x)^p} 
+ (\rho_s x)\, \vev{a_s^p}
$$
It is the first term which dominates for $0<p<1$ and the second for
$1<p$. This reproduces eq.(\ref{fractalburg}).

Another interesting quantity is the probability distribution
function for the gradiant of the velocity. At finite viscocity it is 
formally defined as $P(\zeta)=\vev{\de(\zeta - \d_xu)}$. Its
inviscid limit is the probability distribution function of $\xi(x,t)$
the regular part of $\d_xu(x,t)$ as defined above. As usual, the
stationarity condition leads to a generalized Fokker-Planck equation:
\debut
\zeta P(\zeta) + \d_\zeta(\zeta^2P(\zeta)) + C_1\, \d^2_\zeta P(\zeta)
= \nu \vev{(\d_x^3u)\,\de'(\zeta - \d_xu)}
\label{FPburg}
\fin
with $C_1=-\half C''(0)$. The r.h.s. is the analogue of the dissipative
anomaly but for higher gradiant of the velocity. It is no vanishing due
to the presence of shocks and only shocks contribute to it.

Universality appears in the tails of this probability distribution function
in the inviscid limit.
Consider first the tail for large positive vecolity gradiant, $\zeta\to +\infty$.
It is the regular part of the velocity which dominates in this domain
because the velocity gradiant is large and negative inside shocks.
So for $\zeta\to +\infty$ it is thus legitimate to neglect the
anomalous l.h.s. in the Fokker-Planck equation (\ref{FPburg}).
The differential equation may then be solved and, with appropriate
boundary conditions, it gives \cite{gotoh}:
\debut
P(\zeta)\simeq {\rm const.}\, \zeta\, e^{-\zeta^3/3C_1}
\quad {\rm for}\quad \zeta\to+\infty 
\label{pdf+}
\fin

The behavior of $P(\zeta)$ as $\zeta\to -\infty$ is more tricky 
because shocks contribute significantly in that domain.
One expects a power law behavior:
\debut
P(\zeta)\simeq {\rm const.}\, |\zeta|^{-\al}
\quad {\rm for}\quad \zeta\to-\infty 
\label{pdf-}
\fin
A nice analysis presented in \cite{EvE} shows that
realizability constraints impose $\al<3$ while a finer
analysis of shock formations yields to $\al=7/2$.

It will be interesting to have a statistical description of 
the forced Burgers turbulence as precise as one has for the 
decaying Burgers turbulence, see eg. \cite{burg,kida}.

\vfill
\newpage


\begin{thebibliography}{}
%

\bibitem{temam} R. Temam, {\it Navier-Stokes equations}, 
Studies in mathematics and its apllications, North-Holland, Amsterdam, 1979


\bibitem{yaglom} A.S. Monin and A.M. Yaglom, {\it Statistical Fluid mechanics}, 
vol. 1 and 2, MIT Press, Cambridge MA, 1971 and 1975;\\
G.K. Batchelor, {\it Introduction to Fluid dynamics}, Cambridge Univ. Press, 1967

\bibitem{frisch} U. Frisch, {\it Turbulence: the legacy of A. Kolmogorov},
Cambridge Univ. Press 1995.

\bibitem{K41} A.N. Kolmogorov, C.R. Acad. Sci. URSS 30 (1941) 301.

\bibitem{msr} P.C. Martin, E.D. Siggia and H.A. Rose, Phys. Rev A8 (1973) 423.

\bibitem{fractal} See eg. Parisi G and Frish U. in {\it Turbulence and predictability
in geophysical fluid dynamics}, Intern. School of Phys. E. Fermi, 1983,
eds. M. Ghil, R. Benzi and G. Parisi, North-Holland, Amsterdam, as quoted in [3];\\
R. Benzi, Paladin G., Parisi G. and Vulpiani A., J. Phys. A17 (1984) 3521. 

\bibitem{kraich} R. Kraichnan, Phys. Fluids 10 (1967) 1417.

\bibitem{batch} G.K. Batchelor, Phys. Fluids, Suppl II, 12 (1969) 233.

\bibitem{db99} D. Bernard, Phys. Rev E60 (1999) 6184;\\
E. Lindborg, J. Fluid Mech. 388 (1999) 259.

\bibitem{FaLe} G. Falkovich and V. Lebedev, Phys. Rev. E50 (1994) 3883.

\bibitem{Tab} J. Paret and P. Tabeling, Phys. Fluids, 10 (1998) 3126;\\
J. Paret, M.-C. Jullien and P. Tabeling, Phys. Rev. Lett. 83 (1999) 3418.

\bibitem{also2d} G. Boffetta, A. Celani and M. Vergassola, chao-dyn/9906016;\\
W. B. Daniel and M.A. Rutgers, arXiv:nlin.CD/0005008;\\
E. Lindborg and K. Alvelius, Phys. Fluids 12 (2000) 945.

\bibitem{nurfric} K. Nam, E. Ott, T. Antonsen and P. Guzdar, Phys. Rev. Let 84 (2000) 5134.

\bibitem{fric} D. Bernard, Europhys. Lett. 50 (2000) 333.

\bibitem{Pol} A. Polyakov, Nucl. Phys. B396 (1993) 367.

\bibitem{obuk} A.M. Obukhov, Ivz. Akad. Nauk SSSR Georgr. Geofiz., 13 (1949) 58;\\
S. Corrsin, J. Appl. Phys. 22 (1951) 469.

\bibitem{kraichbis} R. Kraichnan, Phys. Fluids 11 (1968) 945;\\
R. Kraichnan, Phys. Rev. Lett. 72 (19940 1016. 

\bibitem{allscalar} B. Shraimam and E. Siggia, C.R. Acad. Sci. 321 (1995) 279;\\
K. Gawedzki and A. Kupiainen, Phys. Rev. Lett. 75 (1995) 3834;\\
M. Chertkov, G. Falkovich, I. Kolokolov and V. Lebedev, Phys. Rev E52 (1995) 4924.

\bibitem{expo} D. Bernard, K. Gawedzki and A. Kupiainen, Phys. Rev E 54 (1996) 2624.

\bibitem{chertkov} M. Chertkov, Phys. Fluids 10 (1998) 3017.

\bibitem{MazVer} A. Celani, A. Lanotte, A. Mazzino and M. Vergassola, chao-dyn/9909038.

\bibitem{LebBal} E. Balkovsky and V. Lebedev, Phys. Rev E58 (1998) 5776.

\bibitem{plusscalar} On various aspect of the passive scalar problem:\\
 B.I. Shraiman and E.D. Siggia, {\it Scalar turbulence}, preprint (1999);\\
M. Chertkov, G. Falkovich and I. Kolokolov, chao-dyn/9709005;\\
U. Frisch, A. Mazzino, A. Noullez and M. Vergassola, cond-mat/9810074;\\
O. Gat and R. Zeitak, Phys. Rev. E57 (1998)  53331, cond-mat/9711034;\\
I. Arad and I. Procaccia, arXiv:nlin.CD/0001067;\\
On the effect of compressibility:\\
M. Chertkov, I. Kolokolov and M. Vergassola, Phys. Rev. Lett 80 (1998) 512;\\
K. Gawedzki and M. Vergassola, cond-mat/9811399;\\
E. Balkovsky, G. Falkovich and A. Fouxon, chao-dyn/9912027;\\ 
On the vector passive problem:\\
R. Kraichnan and S. Nagarajan, Phys. Fluid 10 (1967) 859;\\
A.P. Kazantsev, Sov. Phys. JETP 26 (1968) 1031;\\
M. Chertkov, G. Falkovich, I. Kolokolov and M. Vergassola, chao-dyn/9906030;

\bibitem{richard} L.F. Richardson, Proc. R. Soc. Lond. A110 (1926) 709.

\bibitem{matheux} Y. LeJan and O. Raimond, Preprint 1998, Orsay.

\bibitem{shrai} B. Shraimam and E. Siggia, Phys. Rev. E77 (1996) 2463.

\bibitem{slow} D. Bernard, K. Gawedzki and A. Kupiainen, J. Stat. Phys. 90 (1998) 519.

\bibitem{lyapu} A. Gamba and  I. Kolokolov, J. Stat. Phys. 85 (1996) 489.

\bibitem{balko} E. Balkovsky and A. Fouxon, Phys. Rev E60 (1999) 4164.

\bibitem{brenier} Y. Brenier, J. Am. Math. Soc.2 (1989) 225;\\
Y. Brenier, Arch. Rat. Mech. Anal. 138 (1997) 319.

\bibitem{shnirel} A. Shnirelman, Commun. Math. Phys. 210 (2000) 541.

\bibitem{burg} J.M. Burgers, {\it the non-linear diffusion equation}, 
D. Reidel Publishing Co, 1974.

\bibitem{kida} S. Kida, J. Fluid. Mech. 93 (1979) 337.

\bibitem{gotoh} T. Gotoh and R. Kraichnan, Phys. Fluids 10 (1998) 2859.

\bibitem{buug}  W.E., K. Khanin, A. Mazel and Y. Sinai, Phys. Rev. Let 78 (1997) 1904.

\bibitem{EvE} W. E. and E. Vanden Eijnden, Phys. Rev. Lett. 83 (1999) 2572;\\
W. E. and E. Vanden Eijnden, chao-dyn/9901006.


%
\end{thebibliography}
\end{document}